\documentclass[aps,pra, twocolumn,superscriptaddress]{revtex4-1}

\usepackage{graphicx}
\usepackage{dcolumn}
\usepackage{bm}
\usepackage[citecolor=blue,colorlinks=true]{hyperref}

\usepackage{caption}
\usepackage{subcaption}

\begin{document}



\title{Rydberg assisted light shift imbalance induced blockade in an atomic ensemble}


\author{Yanfei Tu}
\affiliation{Department of Electrical Engineering and Computer Science, Northwestern University, Evanston, IL 60208, USA}
\author{May E. Kim}
\affiliation{Department of Physics \& Astronomy, Northwestern University, Evanston, IL 60208, USA}
\author{Selim M. Shahriar}
\email[]{shahriar@northwestern.edu}
\affiliation{Department of Electrical Engineering and Computer Science, Northwestern University, Evanston, IL 60208, USA}
\affiliation{Department of Physics \& Astronomy, Northwestern University, Evanston, IL 60208, USA}


\date{\today}

\begin{abstract}
Previously, we had proposed the technique of light shift imbalance induced blockade which leads to a condition where a collection of non-interacting atoms under laser excitation remains combined to a superposition of the ground and the first excited states, thus realizing a collective state quantum bit which in turn can be used to realize a quantum computer. In this paper, we show first that the light shift imbalance by itself is actually not enough to produce such a blockade, and explain the reason why the limitation of our previous analysis had reached this constraint. We then show that by introducing Rydberg interaction, it is possible to achieve such a blockade for a wide range of parameters. Analytic arguments used to establish these results are confirmed by numerical simulations. The fidelity of coupled quantum gates based on such collective state qubits is highly insensitive to the exact number of atoms in the ensemble. As such, this approach may prove to be viable for scalable quantum computing based on neutral atoms.  

\begin{description}
\item[PACS numbers]
32.80.Rm, 32.60.+i, 03.75.Hh, 03.67.Lx
\end{description}
\end{abstract}



\maketitle

\section{Introduction}
   In most protocols for quantum computing or quantum information processing, the fundamental building block is the quantum bit (qubit). A single, neutral atom behaving as a two-level system can be used as a qubit. Compared to ions, neutral atoms have the advantage that they are highly decoupled from electro-magnetic perturbations. However, coupling two qubits using neutral atoms is difficult to achieve. One approach for such coupling makes use of the Rydberg blockade~\cite{PhysRevLett.85.2208,PhysRevLett.87.037901,PhysRevLett.105.193603,PhysRevLett.99.163601,Dudin18052012,PhysRevLett.100.170504,PhysRevLett.97.083003}. In another approach, a cavity mode is used to couple atoms held inside the cavity~\cite{PhysRevLett.75.3788,PhysRevLett.92.127902,PhysRevA.72.032333,PhysRevLett.91.097905}. A key parameter in this approach is the single photon Rabi frequency, which must be much larger than atomic and cavity decay rates. This constraint can only be met by making the cavity very small, which in turn makes it difficult to hold many qubits inside.
 
   One approach for circumventing this constraint is to make use of atomic ensembles. The single photon Rabi frequency for an ensemble scales as  $\sqrt N$, where $N$  is the number of atoms, thus making it possible to make use of a much larger cavity. However, in order to use an ensemble for quantum computing, it is necessary to ensure that it behaves as an effective two-level system.
   
   When exposed to only a single photon (or in a Raman transition, where one leg is exposed to a single photon), an ensemble of two-level atoms does indeed behave like a single two-level system. This property has been used to realize quantum memory elements using such an ensemble~\cite{PhysRevLett.99.260501,Fleischhauer2000395}. However, any protocol that aims to create a two qubit logic gate (such as a CNOT gate) between two ensembles, necessary for realizing a quantum computer, must make use of additional, classical laser fields. Under such excitations, an ensemble no longer behaves like a two-level system. Instead, it exhibits a cascade of energy levels that are equally spaced. When exposed to a classical field, all levels in the cascade get excited~\cite{PhysRev.93.99}, making it impossible to realize a quantum logic gate. In order to overcome this constraint, it is necessary to create conditions under which the cascade is truncated to a two-level system. 
   
   Previously, our group had proposed a scheme for producing such a blockade, using imbalances in light shifts experienced by the collective states~\cite{Shahriar200794,PhysRevA.75.022323}. In that model, the light shifts were calculated by using a perturbation method, keeping terms up to second order in laser intensity. However, it turns out that when the collective excitation is viewed as a product of individual atomic states, an accurate representation for classical laser fields, and in the absence of any interaction between the atoms, the blockade effect disappears. We have verified this conclusion by numerically simulating the evolution of collective states for small values of $N$. It is still possible to produce such a blockade for a laser field described as a superposition of photon number states. However, when the mean photon number in such a field is very large, such as in a classical laser field, the blockade tends to vanish. Thus, in order to produce a blockade under excitation with a classical laser field, we must make use of some interaction between the atoms. In this paper, we propose to make use of interaction induced via excitation to Rydberg states to achieve this goal. 
   
   The rest of the paper is organized as follows. In Section~\ref{sec:collective}, we review briefly the formulation of collective excitation of lambda-type atoms. In Section~\ref{sec:original}, we summarize the model we had developed previously for light shift blockade (LSB) of collective excitation using second order perturbation approximation. In Section~\ref{sec:limitation} we discuss how an alternative formulation of collective excitation allows us to determine the effect of light shift exactly, and identify conditions under which LSB is not possible. In particular, we show that when all excitation fields are classical, there is no blockade. In Section~\ref{sec:twoatom}, we show how the interaction between two Rydberg states can be used to realize LSB even under classical excitation. In section Section~\ref{sec:Natom}, we generalize this process for $N$ atoms and show how LSB works for $N$-atom ensembles. Finally, in Section~\ref{sec:conclusion}, we summarize our results, and present an outlook for using this approach for realizing a multi-qubit quantum computer. 
   
\section{\label{sec:collective}Collective State Model}
In order to avoid the deleterious effect of spontaneous emission, it is useful to realize a qubit based on two states that are long-lived. A convenient example for such a system consists of a Zeeman sublevel in one of the ground hyperfine state (e.g. $m_F=0$, $F=1$, $5^2S_{1/2}$ in $^{87}$Rb) and another Zeeman sublevel in another ground hyperfine state (e.g. $m_F=0$, $F=2$, $5^2S_{1/2}$ in $^{87}$Rb). These levels can be coupled by two laser fields to an intermediate state (e.g. $m_F=1$, $F=2$, $5^2P_{1/2}$ in $^{87}$Rb). When the interaction is highly detuned with respect to the intermediate state, the laser fields cause a Raman transition between the two low lying states, thus producing an effective two-level system. 

\begin{figure}
\includegraphics[width=0.15\textwidth]{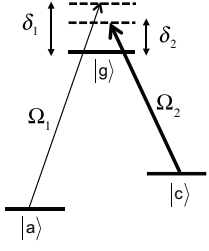}
\caption{Three-level scheme of single atom in an ensemble.}
\label{fig:fig1}
\end{figure}

This is generally known as the $\Lambda$-system, illustrated schematically in Fig.~\ref{fig:fig1}. Here, the two ground states are $\left| a \right\rangle$ and $\left| c \right\rangle $, and the intermediate state is $\left| g \right\rangle$. The states $\left| a \right\rangle$ and $\left| g \right\rangle$ are coupled by a field with a Rabi frequency of $\Omega_1$ and a detuning of $\delta_1$. Likewise, states $\left| c \right\rangle$ and $\left| g \right\rangle$ are coupled by a field with a Rabi frequency of $\Omega_2$ and a detuning of $\delta_2$. In the basis of states $\left| a \right\rangle$, $\left| c \right\rangle$ and $\left| g \right\rangle$, the Hamiltonian under electric dipole and rotating wave approximation, and rotating wave transportation, is given by

\begin{equation}
\widetilde H = \hbar \left[ {\begin{array}{*{20}{c}}
{{\Delta  \mathord{\left/
 {\vphantom {\Delta  2}} \right.
 \kern-\nulldelimiterspace} 2}}&0&{{{{\Omega _1}} \mathord{\left/
 {\vphantom {{{\Omega _1}} 2}} \right.
 \kern-\nulldelimiterspace} 2}}\\
0&{{{ - \Delta } \mathord{\left/
 {\vphantom {{ - \Delta } 2}} \right.
 \kern-\nulldelimiterspace} 2}}&{{{{\Omega _2}} \mathord{\left/
 {\vphantom {{{\Omega _2}} 2}} \right.
 \kern-\nulldelimiterspace} 2}}\\
{{{{\Omega _1}} \mathord{\left/
 {\vphantom {{{\Omega _1}} 2}} \right.
 \kern-\nulldelimiterspace} 2}}&{{{{\Omega _2}} \mathord{\left/
 {\vphantom {{{\Omega _2}} 2}} \right.
 \kern-\nulldelimiterspace} 2}}&{ - \delta }
\end{array}} \right],
\label{eq:eqn1}
\end{equation}
where $\delta \equiv({\delta_1}+{\delta_2})/2$ is the average detuning and $\Delta \equiv({\delta_1}-{\delta_2})$ is the two-photon detuning. In what follows, we will assume that $\delta$ is very large compared to $\Omega_1$ and $\Omega_2$, as well as the decay rate, $\Gamma$, of the state $\left| g \right\rangle$. We will further assume that the two lasers are co-propagating.

For $N$ such non-interacting atoms, the ensemble can be modeled using symmetric collective states, also known as symmetric Dicke states~\cite{PhysRev.93.99}. The first few states are defined as follows:
\begin{equation}
\begin{array}{l}
\left| A \right\rangle  \equiv \left| {{a_1},{a_2}, \cdot  \cdot ,{a_N}} \right\rangle ,\\
\left| {{G_1}} \right\rangle  \equiv \frac{1}{{\sqrt N }}\sum\limits_{j = 1}^N {\left| {{a_1},{a_2}, \cdot  \cdot ,{g_j}, \cdot  \cdot ,{a_N}} \right\rangle } ,\\
\left| {{C_1}} \right\rangle  \equiv \frac{1}{{\sqrt N }}\sum\limits_{j = 1}^N {\left| {{a_1},{a_2}, \cdot  \cdot ,{c_j}, \cdot  \cdot ,{a_N}} \right\rangle } ,\\
\left| {{G_2}} \right\rangle  \equiv \frac{1}{{\sqrt {{}^N{C_2}} }}\sum\limits_{j,k(j \ne k)}^{{}^N{C_2}} {\left| {{a_1},{a_2}, \cdot  \cdot ,{g_j}, \cdot  \cdot ,{g_k}, \cdot  \cdot ,{a_N}} \right\rangle } ,\\
\left| {{C_2}} \right\rangle  \equiv \frac{1}{{\sqrt {{}^N{C_2}} }}\sum\limits_{j,k(j \ne k)}^{{}^N{C_2}} {\left| {{a_1},{a_2}, \cdot  \cdot ,{c_j}, \cdot  \cdot ,{c_k}, \cdot  \cdot ,{a_N}} \right\rangle } ,\\
\left| {{G_{1,1}}} \right\rangle  \equiv \frac{1}{{\sqrt {2{}^N{C_2}} }}\sum\limits_{j,k(j \ne k)}^{2{}^N{C_2}} {\left| {{a_1},{a_2}, \cdot  \cdot ,{g_j}, \cdot  \cdot ,{c_k}, \cdot  \cdot ,{a_N}} \right\rangle } ,\\
\left| {{G_{2,1}}} \right\rangle  \equiv \frac{1}{{\sqrt {3{}^N{C_3}} }}\sum\limits_{j,k,l(j \ne k \ne l)}^{3{}^N{C_3}} {\left| {{a_1},{a_2}, \cdot  \cdot ,{g_j}, \cdot  \cdot ,{g_k}, \cdot  \cdot ,{c_l}, \cdot  \cdot ,{a_N}} \right\rangle } ,\\
\left| {{G_{1,2}}} \right\rangle  \equiv \frac{1}{{\sqrt {3{}^N{C_3}} }}\sum\limits_{j,k,l(j \ne k \ne l)}^{3{}^N{C_3}} {\left| {{a_1},{a_2}, \cdot  \cdot ,{g_j}, \cdot  \cdot ,{c_k}, \cdot  \cdot ,{c_l}, \cdot  \cdot ,{a_N}} \right\rangle } .
\end{array}
\label{eq:eqn2}
\end{equation}
where $^N{C_M} \equiv \left( {\begin{array}{*{20}{c}}N\\M\end{array}} \right) \equiv {{N!} \mathord{\left/{\vphantom {{N!} {\left[ {M!\left( {N - M} \right)!} \right]}}} \right.
 \kern-\nulldelimiterspace} {\left[ {M!\left( {N - M} \right)!} \right]}}$.

\begin{figure}
\includegraphics[width=0.4\textwidth]{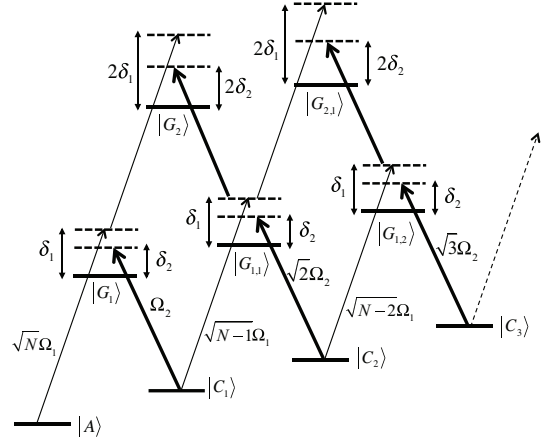}
\caption{Schematic illustration of the relevant collective states and the corresponding coupling rates.}
\label{fig:fig2}
\end{figure}

In Ref~\cite{resham}, we have shown that the system remains confined to a generalized form of these symmetric collective states, independent of the relative separation between the atoms (and hence the size of the ensemble), as long as it is assumed that each atom sees the same amplitude of the Rabi frequency, and the same laser frequency (i.e., any residual Doppler shift of the Raman transition frequency due to the motion of the atoms is negligible). The generalized form of the symmetric states are formally the same as those in Eqn.~(\ref{eq:eqn2}), except that the excited states incorporate the relevant spatial phases of the fields at the location of a given atom. This can be understood by noting that any phase factors accompanying the Rabi frequencies in the Hamiltonian of Eqn.~(\ref{eq:eqn1}) can be transformed out to produce a version of the Hamiltonian where the Rabi frequencies are real. The transformation necessary for this transfers the phases to the basis states. We refer the reader to Ref~\cite{resham} for details.

The collective states of Eqn.~(\ref{eq:eqn2}) are illustrated schematically in Fig.~\ref{fig:fig2}. Here, for example, $\left| {{G_1}} \right\rangle$ represents a state where only one atom on average is excited to state $\left| g \right\rangle$, with the rest remaining in state $\left| a \right\rangle$. Similarly, $\left| {{C_1}} \right\rangle$ represents a state where only one atom on average is excited to state $\left| c \right\rangle$, with the rest remaining in state $\left| a \right\rangle$, and so on. In our blockade scheme, we try to confine the system to the two lowest energy states $\left| A \right\rangle$ and $\left| {{C_1}} \right\rangle$. If we could achieve this and minimize the excitations to the first few higher energy states, then excitations to even higher states will be almost nonexistent. It can be shown that the total number of symmetric states is ${N_S} = {{\left( {N + 2} \right)!} \mathord{\left/{\vphantom {{\left( {N + 2} \right)!} {2N!}}} \right.
 \kern-\nulldelimiterspace} {2N!}}$. For large $N$, ${N_S} = {{{N^2}} \mathord{\left/
 {\vphantom {{{N^2}} 2}} \right.\kern-\nulldelimiterspace} 2}$ so that the size of the Hamiltonian scales as $N^4$. Thus, an analysis of the evolution of the complete system exactly in this picture is computationally intractable. However, a plausible way to explore the possibility of finding the condition for the blockade is to truncate the system to a small size, and show that the excitation to the excluded states are negligible.
   
   Here, we choose to truncate the system to six levels: $\left| A \right\rangle$, $\left| {{G_1}} \right\rangle$, $\left| {{C_1}} \right\rangle$, $\left| {{G_{1,1}}} \right\rangle$, $\left| {{C_2}} \right\rangle$ and $\left| {{G_{1,2}}} \right\rangle$. If the condition we find for the blockade shows negligible excitation to states that have non-zero coupling to the excluded states, the truncation would then be justified. The Hamiltonian for these states can be expressed as~\cite{Shahriar200794}
\begin{widetext}   
\begin{equation}  
H = \hbar \left[ {\begin{array}{*{20}{c}}
{{\Delta  \mathord{\left/
 {\vphantom {\Delta  2}} \right.
 \kern-\nulldelimiterspace} 2}}&
{{{\sqrt {N-2} {\Omega _1}} \mathord{\left/
 {\vphantom {{\sqrt {N-2} {\Omega _1}} 2}} \right.
 \kern-\nulldelimiterspace} 2}}
&0&0&0&0\\
{{{\sqrt N {\Omega _1}} \mathord{\left/
 {\vphantom {{\sqrt N {\Omega _1}} 2}} \right.
 \kern-\nulldelimiterspace} 2}}&{ - \delta }&{{{{\Omega _2}} \mathord{\left/
 {\vphantom {{{\Omega _2}} 2}} \right.
 \kern-\nulldelimiterspace} 2}}&0&0&0\\
0&{{{{\Omega _2}} \mathord{\left/
 {\vphantom {{{\Omega _2}} 2}} \right.
 \kern-\nulldelimiterspace} 2}}&{ - {\Delta  \mathord{\left/
 {\vphantom {\Delta  2}} \right.
 \kern-\nulldelimiterspace} 2}}&{{{\sqrt {N - 1} {\Omega _1}} \mathord{\left/
 {\vphantom {{\sqrt {N - 1} {\Omega _1}} 2}} \right.
 \kern-\nulldelimiterspace} 2}}&0&0\\
0&0&{{{\sqrt {N - 1} {\Omega _1}} \mathord{\left/
 {\vphantom {{\sqrt {N - 1} {\Omega _1}} 2}} \right.
 \kern-\nulldelimiterspace} 2}}&{ - \left( {\delta  + \Delta } \right)}&{{{\sqrt 2 {\Omega _2}} \mathord{\left/
 {\vphantom {{\sqrt 2 {\Omega _2}} 2}} \right.
 \kern-\nulldelimiterspace} 2}}&0\\
0&0&0&{{{\sqrt 2 {\Omega _2}} \mathord{\left/
 {\vphantom {{\sqrt 2 {\Omega _2}} 2}} \right.
 \kern-\nulldelimiterspace} 2}}&{ - {{3\Delta } \mathord{\left/
 {\vphantom {{3\Delta } 2}} \right.
 \kern-\nulldelimiterspace} 2}}&{{{\sqrt {N - 2} {\Omega _1}} \mathord{\left/
 {\vphantom {{\sqrt {N - 2} {\Omega _1}} 2}} \right.
 \kern-\nulldelimiterspace} 2}}\\
0&0&0&0&{{{\sqrt {N - 2} {\Omega _1}} \mathord{\left/
 {\vphantom {{\sqrt {N - 2} {\Omega _1}} 2}} \right.
 \kern-\nulldelimiterspace} 2}}&{ - \left( {\delta  + 2\Delta } \right)}
\end{array}} \right].
\label{eq:eqn3}
\end{equation}   
\end{widetext}
   
\section{\label{sec:original}Original Model for Light Shift Blockade}
The Hamiltonian in Eqn.~(\ref{eq:eqn3}) can be further simplified by adiabatically eliminating the states $\left| {{G_1}} \right\rangle$, $\left| {{G_{1,1}}} \right\rangle$, and $\left| {{G_{1,2}}} \right\rangle$ when $\delta \gg \sqrt N {\Omega _1}$, $\Omega_2$, $\Delta$, and $N \gg 1$. The reduced Hamiltonian in the basis of states $\left| A \right\rangle$, $\left| {{C_1}} \right\rangle$ and $\left| {{C_2}} \right\rangle$ is
\begin{equation}  
\widetilde H = \hbar \left[ {\begin{array}{*{20}{c}}
{{\varepsilon _A} + {\Delta  \mathord{\left/
 {\vphantom {\Delta  2}} \right.
 \kern-\nulldelimiterspace} 2}}&{{\Omega  \mathord{\left/
 {\vphantom {\Omega  2}} \right.
 \kern-\nulldelimiterspace} 2}}&0\\
{{\Omega  \mathord{\left/
 {\vphantom {\Omega  2}} \right.
 \kern-\nulldelimiterspace} 2}}&{{\varepsilon _{C1}} - {\Delta  \mathord{\left/
 {\vphantom {\Delta  2}} \right.
 \kern-\nulldelimiterspace} 2}}&{\sqrt {\frac{{2\left( {N - 1} \right)}}{N}} {\Omega  \mathord{\left/
 {\vphantom {\Omega  2}} \right.
 \kern-\nulldelimiterspace} 2}}\\
0&{\sqrt {\frac{{2\left( {N - 1} \right)}}{N}} {\Omega  \mathord{\left/
 {\vphantom {\Omega  2}} \right.
 \kern-\nulldelimiterspace} 2}}&{{\varepsilon _{C2}} - {{3\Delta } \mathord{\left/
 {\vphantom {{3\Delta } 2}} \right.
 \kern-\nulldelimiterspace} 2}}
\end{array}} \right],
\label{eq:eqn4}
\end{equation}  
where ${\varepsilon _A} ={N\Omega _1^2}/{4\delta}$, ${\varepsilon _{C1}} = {{\left[ {\Omega _2^2 + \left( {N - 1} \right)\Omega _1^2} \right]} \mathord{\left/
 {\vphantom {{\left[ {\Omega _2^2 + \left( {N - 1} \right)\Omega _1^2} \right]} {4\delta }}} \right.
 \kern-\nulldelimiterspace} {4\delta }}$, and ${\varepsilon _{C2}} = {{\left[ {2\Omega _2^2 + \left( {N - 2} \right)\Omega _1^2} \right]} \mathord{\left/{\vphantom {{\left[ {2\Omega _2^2 + \left( {N - 2} \right)\Omega _1^2} \right]} {4\delta }}} \right.\kern-\nulldelimiterspace} {4\delta }}$ are the lowest order light-shifts of the states $\left| A \right\rangle$, $\left| {{C_1}} \right\rangle$ and $\left| {{C_2}} \right\rangle$ respectively, and $\Omega  \equiv {{\sqrt N {\Omega _1}{\Omega _2}} \mathord{\left/{\vphantom {{\sqrt N {\Omega _1}{\Omega _2}} {2\delta }}} \right.
 \kern-\nulldelimiterspace} {2\delta }}$ is the Raman Rabi frequency. We can work out the LSB conditions with this Hamiltonian. By making the light shifts in the states $\left| A \right\rangle$ and $\left| {{C_1}} \right\rangle$ equal and the shift in $\left| {{C_2}} \right\rangle$ highly detuned from them, we can eliminate the excitation to $\left| {{C_2}} \right\rangle$.

The states $\left| A \right\rangle$ and $\left| {{C_1}} \right\rangle$ are resonant when $\Delta  = {\varepsilon _{C1}} - {\varepsilon _A} \approx {{\left( {\Omega _2^2 - \Omega _1^2} \right)} \mathord{\left/ {\vphantom {{\left( {\Omega _2^2 - \Omega _1^2} \right)} {4\delta }}} \right.
 \kern-\nulldelimiterspace} {4\delta }}$. Upon subtraction of a suitably chosen term $\left( {{\varepsilon _A} + {\Delta  \mathord{\left/ {\vphantom {\Delta  2}} \right. \kern-\nulldelimiterspace} 2}} \right)$  from the diagonal term in the Hamiltonian and the approximation that $N \gg 1$, we get
\begin{equation} 
\widetilde H = \hbar \left[ {\begin{array}{*{20}{c}}
0&{{\Omega  \mathord{\left/
 {\vphantom {\Omega  2}} \right.
 \kern-\nulldelimiterspace} 2}}&0\\
{{\Omega  \mathord{\left/
 {\vphantom {\Omega  2}} \right.
 \kern-\nulldelimiterspace} 2}}&0&{{\Omega  \mathord{\left/
 {\vphantom {\Omega  {\sqrt 2 }}} \right.
 \kern-\nulldelimiterspace} {\sqrt 2 }}}\\
0&{{\Omega  \mathord{\left/
 {\vphantom {\Omega  {\sqrt 2 }}} \right.
 \kern-\nulldelimiterspace} {\sqrt 2 }}}&{{\Delta _B}}
\end{array}} \right],
\label{eq:eqn5}
\end{equation}
where the blockade shift is defined as ${\Delta _B} \equiv \left( {{\varepsilon _{C2}} - {\varepsilon _{C1}}} \right) - \left( {{\varepsilon _{C1}} - {\varepsilon _A}} \right)$. This quantity vanishes for the first order values of the light shifts ${\varepsilon _A}$, ${\varepsilon _{C1}}$, and ${\varepsilon _{C2}}$ shown above, so that there is no blockade effect. However, to second order approximation, the blockade shift is ${\Delta _B} =  - {{\left( {\Omega _1^4 + \Omega _2^4} \right)} \mathord{\left/ {\vphantom {{\left( {\Omega _1^4 + \Omega _2^4} \right)} {\left( {8{\delta ^3}} \right)}}} \right. \kern-\nulldelimiterspace} {\left( {8{\delta ^3}} \right)}}$. If we operate under condition where ${\Delta _B} \gg {\Omega  \mathord{\left/ {\vphantom {\Omega  {\sqrt 2 }}} \right. \kern-\nulldelimiterspace} {\sqrt 2 }}$, the transition to $\left| {{C_2}} \right\rangle$ becomes inconsequentially small and the ensemble of atoms oscillates between the collective states $\left| A \right\rangle$ and $\left| {{C_1}} \right\rangle$.

\begin{figure}
\includegraphics[width=0.35\textwidth]{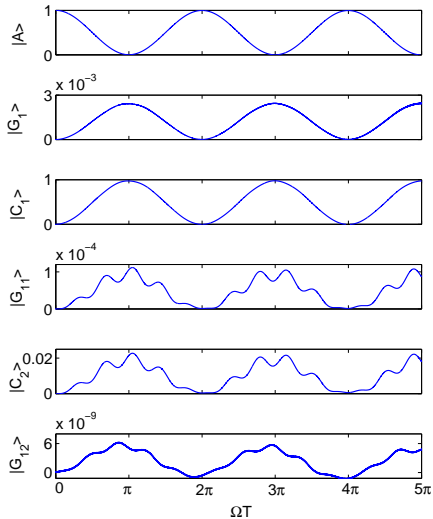}
\caption{ Exact numerical solution of the evolution of the states using the LSB parameters (in units of $\Gamma$): $\Omega_1=0.001$, $\Omega_2 = 100$, $N = 2500$, $\delta = 1000$ and $\Delta = 2.497$. The plot is for 5$\pi$ oscillations. The vertical axis is the population of the indicated collective state.}
\label{fig:fig3}
\end{figure}

   We have also determined numerically, for $N = 2500$, the evolution of the population for the six collective states in the truncated system, using the Hamiltonian of Eqn.~(\ref{eq:eqn3}), without resorting to adiabatic elimination. The results are illustrated in Fig.~\ref{fig:fig3}, for a set of parameters that satisfy the LSB condition identified above. As can be seen from this figure, nearly all the population stays between levels $\left| A \right\rangle$ and $\left| {{C_1}} \right\rangle$, undergoing Rabi oscillations between them. The residual excitations of the other four states are very small, and can be made smaller by using weaker Rabi frequencies. Note that we have ignored the decay of the $\left| g \right\rangle$ states (at the rate of $\Gamma$), which is a valid approximation for $\delta  \gg \Gamma$. 

\section{\label{sec:limitation}Limitations of the Original Model for Light Shift Blockade}
\begin{figure*}
        \centering
        \begin{subfigure}[b]{0.35\textwidth}
                \caption{}
                \includegraphics[width=\textwidth]{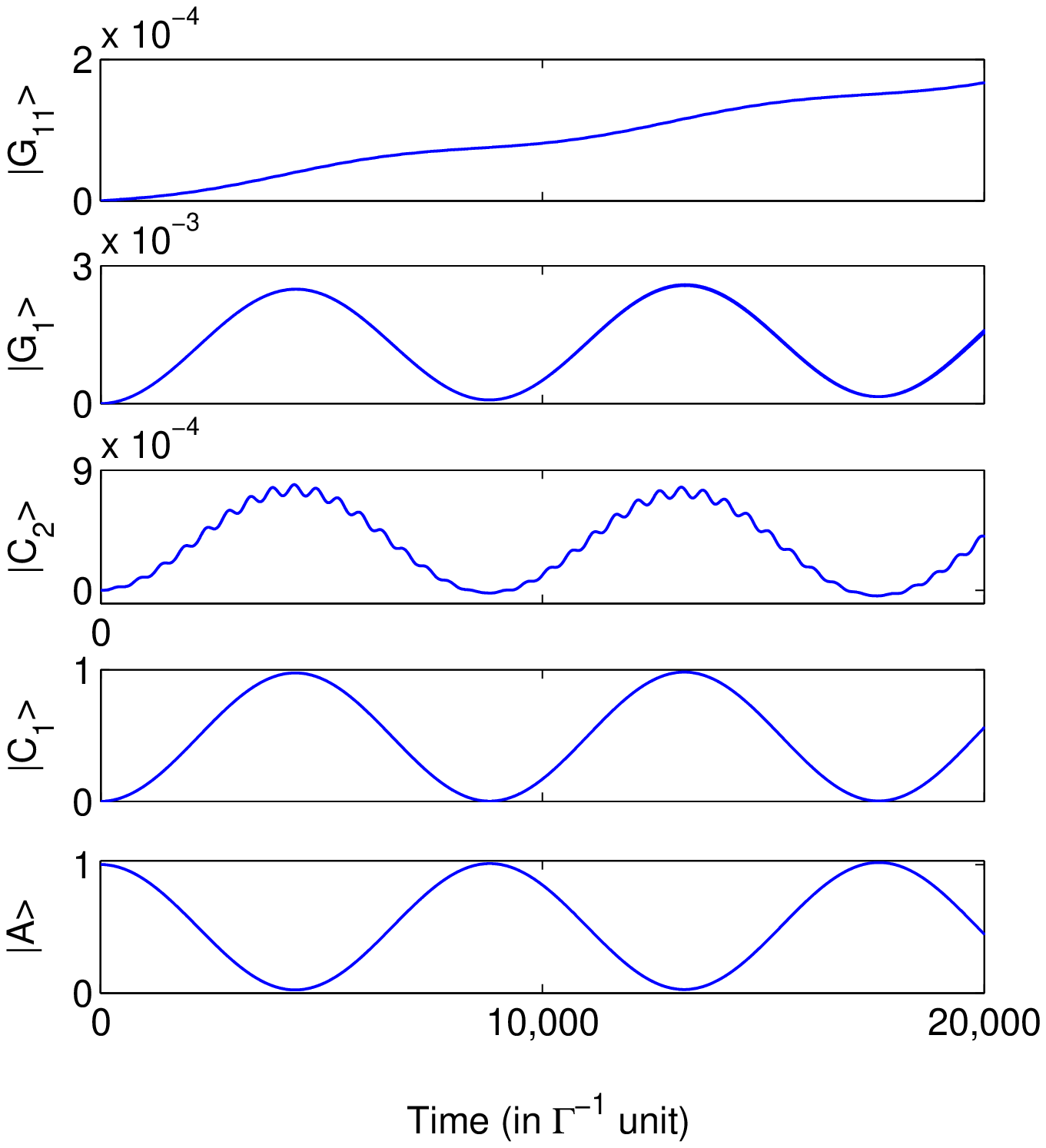}      
                \label{fig:fig4a}
        \end{subfigure}
        \qquad
        \begin{subfigure}[b]{0.35\textwidth}
                \caption{}
                \includegraphics[width=\textwidth]{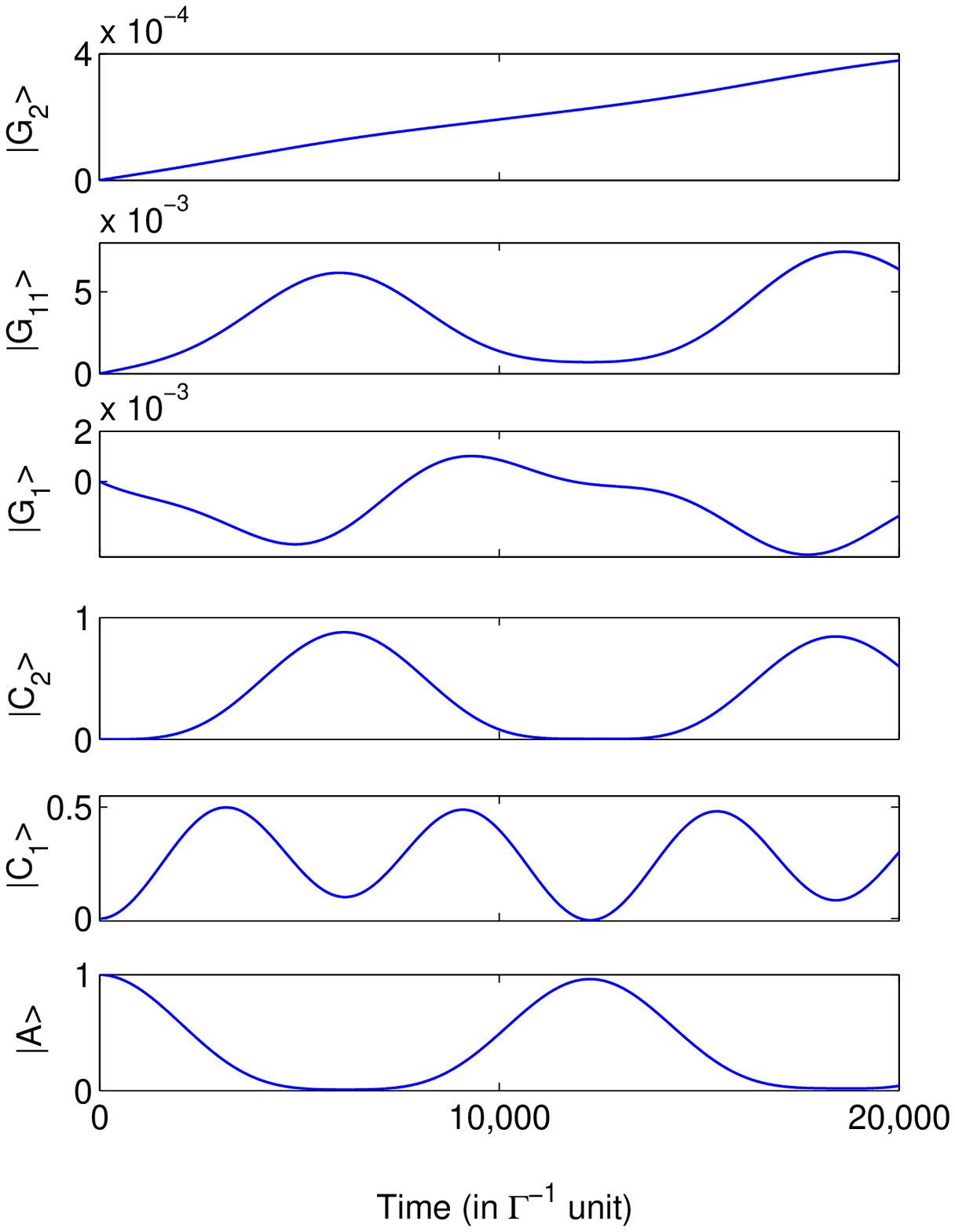}                
                \label{fig:fig4b}
        \end{subfigure}
\caption{Numerical solution of the evolution of the collective states of two atoms. Here, $\Omega_1=0.001$, $\Omega_2 = 100$, $N = 2500$, $\delta = 1000$ and $\Delta = 2.497$. (in units of $\Gamma$). (a): Collective states of two atoms when $\left| {{G_2}} \right\rangle$ is eliminated. (b): Collective states of two atoms with the full Hamiltonian.}
\label{fig:fig4}
\end{figure*}

In the preceding section, we showed that the numerical simulation of the truncated system appears to validate the LSB process. For a large value of $N$, this result is still an approximation. However, the system can be modeled exactly for very small values of $N$. In particular, if we choose $N = 2$, there are only 6 collective states altogether. Thus, it is possible to check without truncation whether the LSB process holds in this case. Referring back to Fig.~\ref{fig:fig2}, the complete set of collective states for $N = 2$ consists of $\left| A \right\rangle$, $\left| {{G_1}} \right\rangle$, $\left| {{C_1}} \right\rangle$, $\left| {{G_{1,1}}} \right\rangle$, $\left| {{C_2}} \right\rangle$ and $\left| {{G_2}} \right\rangle$. We determined the evolution of this system numerically, starting with the system being in the $\left| A \right\rangle$ state. The results are illustrated in Fig.~\ref{fig:fig4}. In Fig.~\ref{fig:fig4a}, we show the population of the collective states under the approximation that the state $\left| {{G_2}} \right\rangle$ can be neglected completely, since $\delta  \gg \sqrt N {\Omega _1}$ and $\delta  \gg {\Omega _2}$, corresponding to very small populations in states $\left| {{G_1}} \right\rangle$ and $\left| {{G_{1,1}}} \right\rangle$. As can be seen, the result is consistent with LSB, since the maximum population of $\left| {{C_2}} \right\rangle$ is very small. In Fig.~\ref{fig:fig4b}, we relax this approximation, and keep the state $\left| {{G_2}} \right\rangle$ in the system. This produces an apparently surprising result. The population in $\left| {{C_2}} \right\rangle$ can now reach almost unity for some interaction time. Thus, the LSB process is strongly violated. It should be noted that the maximum population of $\left| {{G_2}} \right\rangle$ is negligible (Fig.~\ref{fig:fig4a}), so that ignoring the excitation to $\left| {{G_2}} \right\rangle$ seems to be a reasonable one. Yet, the relaxation of this approximation modifies the population dynamics in a very significant way.

In order to understand this behavior, it is instructive first to consider the process of collective excitation more explicitly. Specially, it can be shown that, for excitation by semi-classical fields, and in the absence of interaction between the atoms, the general quantum state of an ensemble is always given by the outer (tensor) product of the quantum states of the individual atoms~\cite{PhysRevA.6.2211}. The collective states representation of the evolution of such a system is merely an alternative way of describing the process. To illustrate this explicitly, let us consider a case involving two-level atoms, with $\left| a \right\rangle$ and $\left| c \right\rangle$ being the lower and higher energy levels respectively.

Let us denote by $\left| {{\psi _i}} \right\rangle$ the quantum state of the $i$-th atom. Then, the total quantum state of the system, $\left| \Psi  \right\rangle$, is given by: $\left| \Psi  \right\rangle  = \mathop \Pi \limits_{i = 1}^N \left| {{\psi _i}} \right\rangle$. Thus, if we write $\left| {{\psi _i}} \right\rangle  = {\alpha _i}\left| {{a_i}} \right\rangle  + {\beta _i}\left| {{c_i}} \right\rangle $, then $\left| \Psi  \right\rangle  = \mathop \Pi \limits_{i = 1}^N \left( {{\alpha _i}\left| {{a_i}} \right\rangle  + {\beta _i}\left| {{c_i}} \right\rangle } \right)$. For simplicity, let us assume that $N=2$. We then get: $\left| \Psi  \right\rangle  = \left( {{\alpha _1}\left| {{a_1}} \right\rangle  + {\beta _1}\left| {{c_1}} \right\rangle } \right)\left( {{\alpha _2}\left| {{a_2}} \right\rangle  + {\beta _2}\left| {{c_2}} \right\rangle } \right)$. Consider the product state basis which is spanned by $\left| {{a_1}{a_2}} \right\rangle$, $\left| {{a_1}{c_2}} \right\rangle$, $\left| {{c_1}{a_2}} \right\rangle$ and $\left| {{c_1}{c_2}} \right\rangle$. The total state can thus be written as

\begin{equation}
\begin{array}{l}
\left| \Psi  \right\rangle  = {\alpha _1}{\alpha _2}\left| {{a_1}{a_2}} \right\rangle  + {\alpha _1}{\beta _2}\left| {{a_1}{c_2}} \right\rangle  + {\beta _1}{\alpha _2}\left| {{c_1}{a_2}} \right\rangle \\\qquad + {\beta _1}{\beta _2}\left| {{c_1}{c_2}} \right\rangle  = \left[ {\begin{array}{*{20}{c}}
{{\alpha _1}{\alpha _2}}\\
{{\alpha _1}{\beta _2}}\\
{{\beta _1}{\alpha _2}}\\
{{\beta _1}{\beta _2}}
\end{array}} \right].
\end{array}
\label{eq:eqn6}
\end{equation}

Consider next the complete collective state basis spanned by $\left| {{a_1}{a_2}} \right\rangle$, $\left|  +  \right\rangle  = {{\left( {\left| {{a_1}{c_2}} \right\rangle  + \left| {{c_1}{a_2}} \right\rangle } \right)} \mathord{\left/ {\vphantom {{\left( {\left| {{a_1}{c_2}} \right\rangle  + \left| {{c_1}{a_2}} \right\rangle } \right)} {\sqrt 2 }}} \right. \kern-\nulldelimiterspace} {\sqrt 2 }}$, $\left|  -  \right\rangle  = {{\left( {\left| {{a_1}{c_2}} \right\rangle  - \left| {{c_1}{a_2}} \right\rangle } \right)} \mathord{\left/ {\vphantom {{\left( {\left| {{a_1}{c_2}} \right\rangle  - \left| {{c_1}{a_2}} \right\rangle } \right)} {\sqrt 2 }}} \right. \kern-\nulldelimiterspace} {\sqrt 2 }}$, and $\left| {{c_1}{c_2}} \right\rangle$. This basis is simply related to the product state basis by a 45° rotation in the plane of $\left| {{a_1}{c_2}} \right\rangle$ and $\left| {{c_1}{a_2}} \right\rangle$, so that the rotation matrix can be written as
\begin{equation}
R = \left[ {\begin{array}{*{20}{c}}
1&0&0&0\\
0&{\frac{1}{{\sqrt 2 }}}&{\frac{1}{{\sqrt 2 }}}&0\\
0&{\frac{1}{{\sqrt 2 }}}&{ - \frac{1}{{\sqrt 2 }}}&0\\
0&0&0&1
\end{array}} \right].
\label{eq:eqn7}
\end{equation}
Thus, the total state in the collective state basis can be written as
\begin{equation}
{\left| \Psi  \right\rangle _c} = R\left| \Psi  \right\rangle  = \left[ {\begin{array}{*{20}{c}}
{{\alpha _1}{\alpha _2}}\\
{{{\left( {{\alpha _1}{\beta _2} + {\beta _1}{\alpha _2}} \right)} \mathord{\left/
 {\vphantom {{\left( {{\alpha _1}{\beta _2} + {\beta _1}{\alpha _2}} \right)} {\sqrt 2 }}} \right.
 \kern-\nulldelimiterspace} {\sqrt 2 }}}\\
{{{\left( {{\alpha _1}{\beta _2} - {\beta _1}{\alpha _2}} \right)} \mathord{\left/
 {\vphantom {{\left( {{\alpha _1}{\beta _2} - {\beta _1}{\alpha _2}} \right)} {\sqrt 2 }}} \right.
 \kern-\nulldelimiterspace} {\sqrt 2 }}}\\
{{\beta _1}{\beta _2}}
\end{array}} \right].
\label{eq:eqn8}
\end{equation}
Similarly, we can represent the Hamiltonian in these different bases. In the rotating wave picture, the Hamiltonian for a single atom can be expressed as
\begin{equation}
{H_1} = \hbar \left[ {\begin{array}{*{20}{c}}
0&{{\Omega  \mathord{\left/
 {\vphantom {\Omega  2}} \right.
 \kern-\nulldelimiterspace} 2}}\\
{{\Omega  \mathord{\left/
 {\vphantom {\Omega  2}} \right.
 \kern-\nulldelimiterspace} 2}}&{ - \delta }
\end{array}} \right],
\label{eq:eqn9}
\end{equation}
where $\Omega$ is the Rabi frequency and $\delta  = \omega  - \left( {{\omega _c} - {\omega _a}} \right)$ is the detuning of the laser frequency from the resonance frequency of the two states. When there are two atoms, the Hamiltonian in the basis of states $\left| {{a_1}{a_2}} \right\rangle$, $\left| {{a_1}{c_2}} \right\rangle$, $\left| {{c_1}{a_2}} \right\rangle$ and $\left| {{c_1}{c_2}} \right\rangle$ is $H = {H_1} \otimes {I_2} + {I_1} \otimes {H_2}$ where $I_i$ is the identity matrix and $H_i$ is the Hamiltonian for the $i$-th atom. For example,

\[\begin{array}{*{20}{l}}
\left\langle {{a_1}{a_2}} \right|H\left| {{c_1}{a_2}} \right\rangle \\
 = \left\langle {{a_1}{a_2}} \right|\left( {{H_1} \otimes {I_2}} \right)\left| {{c_1}{a_2}} \right\rangle  + \left\langle {{a_1}{a_2}} \right|\left( {{I_1} \otimes {H_2}} \right)\left| {{c_1}{a_2}} \right\rangle 
\\
{ = \left\langle {{a_1}} \right|{H_1}\left| {{c_1}} \right\rangle \left\langle {{a_2}} \right|{H_1}\left| {{a_2}} \right\rangle  + \left\langle {{a_1}} \right|{I_1}\left| {{c_1}} \right\rangle \left\langle {{a_2}} \right|{H_2}\left| {{a_2}} \right\rangle }\\
 = \left\langle {{a_1}} \right|{H_1}\left| {{c_1}} \right\rangle \\
 = {{{\Omega _1}} \mathord{\left/
 {\vphantom {{{\Omega _1}} 2}} \right.
 \kern-\nulldelimiterspace} 2}
\end{array}\]
Thus, the Hamiltonian can be written as
\begin{equation}
H = \hbar \left[ {\begin{array}{*{20}{c}}
0&{{{{\Omega _2}} \mathord{\left/
 {\vphantom {{{\Omega _2}} 2}} \right.
 \kern-\nulldelimiterspace} 2}}&{{{{\Omega _1}} \mathord{\left/
 {\vphantom {{{\Omega _1}} 2}} \right.
 \kern-\nulldelimiterspace} 2}}&0\\
{{{{\Omega _2}} \mathord{\left/
 {\vphantom {{{\Omega _2}} 2}} \right.
 \kern-\nulldelimiterspace} 2}}&{ - \delta }&0&{{{{\Omega _1}} \mathord{\left/
 {\vphantom {{{\Omega _1}} 2}} \right.
 \kern-\nulldelimiterspace} 2}}\\
{{{{\Omega _1}} \mathord{\left/
 {\vphantom {{{\Omega _1}} 2}} \right.
 \kern-\nulldelimiterspace} 2}}&0&{ - \delta }&{{{{\Omega _2}} \mathord{\left/
 {\vphantom {{{\Omega _2}} 2}} \right.
 \kern-\nulldelimiterspace} 2}}\\
0&{{{{\Omega _1}} \mathord{\left/
 {\vphantom {{{\Omega _1}} 2}} \right.
 \kern-\nulldelimiterspace} 2}}&{{{{\Omega _2}} \mathord{\left/
 {\vphantom {{{\Omega _2}} 2}} \right.
 \kern-\nulldelimiterspace} 2}}&{ - 2\delta }
\end{array}} \right],
\label{eq:eqn10}
\end{equation}
where the Rabi frequencies are assumed to be real. Under a 45$^\circ $  rotation in the plane of $\left| {{a_1}{c_2}} \right\rangle$ and $\left| {{c_1}{a_2}} \right\rangle$, the new Hamiltonian in the basis $\left| {{a_1}{a_2}} \right\rangle$, $\left|  +  \right\rangle$, $\left|  -  \right\rangle$, $\left| {{c_1}{c_2}} \right\rangle$ is
\begin{equation}
\begin{array}{l}
H' = {R^{ - 1}}HR\\
\quad \;\;= \hbar \left[ {\begin{array}{*{20}{c}}
0&{\frac{{{\Omega _1} + {\Omega _2}}}{{2\sqrt 2 }}}&{ - \frac{{{\Omega _1} - {\Omega _2}}}{{2\sqrt 2 }}}&0\\
{\frac{{{\Omega _1} + {\Omega _2}}}{{2\sqrt 2 }}}&{ - \delta }&0&{\frac{{{\Omega _1} + {\Omega _2}}}{{2\sqrt 2 }}}\\
{ - \frac{{{\Omega _1} - {\Omega _2}}}{{2\sqrt 2 }}}&0&{ - \delta }&{\frac{{{\Omega _1} - {\Omega _2}}}{{2\sqrt 2 }}}\\
0&{\frac{{{\Omega _1} + {\Omega _2}}}{{2\sqrt 2 }}}&{ - \frac{{{\Omega _1} - {\Omega _2}}}{{2\sqrt 2 }}}&{ - 2\delta }
\end{array}} \right]
\end{array}.
\label{eq:eqn11}
\end{equation}

For $\Omega  = {\Omega _1} = {\Omega _2}$, the asymmetric state, $\left|  -  \right\rangle$, is decoupled from the other states, and the Hamiltonian becomes
\begin{equation}
H' = \hbar \left[ {\begin{array}{*{20}{c}}
0&{{{\sqrt 2 \Omega } \mathord{\left/
 {\vphantom {{\sqrt 2 \Omega } 2}} \right.
 \kern-\nulldelimiterspace} 2}}&0&0\\
{{{\sqrt 2 \Omega } \mathord{\left/
 {\vphantom {{\sqrt 2 \Omega } 2}} \right.
 \kern-\nulldelimiterspace} 2}}&{ - \delta }&0&{{{\sqrt 2 \Omega } \mathord{\left/
 {\vphantom {{\sqrt 2 \Omega } 2}} \right.
 \kern-\nulldelimiterspace} 2}}\\
0&0&{ - \delta }&0\\
0&{{{\sqrt 2 \Omega } \mathord{\left/
 {\vphantom {{\sqrt 2 \Omega } 2}} \right.
 \kern-\nulldelimiterspace} 2}}&0&{ - 2\delta }
\end{array}} \right].
\label{eq:eqn12}
\end{equation}
The Hamiltonian in Eqn.~(\ref{eq:eqn12}) describes the situation where only symmetric collective states are excited.

   This is also evident by noting that the general collective state can now be expressed as ${\left| \Psi  \right\rangle _c} = {\alpha ^2}\left| {aa} \right\rangle  + \sqrt 2 \alpha \beta \left|  +  \right\rangle  + {\beta ^2}\left| {cc} \right\rangle$, where $\alpha  = {\alpha _1} = {\alpha _2}$ and $\beta  = {\beta _1} = {\beta _2}$ (since ${\Omega _1} = {\Omega _2}$). The form of this state shows clearly that it is impossible to suppress excitation to the $\left|  cc  \right\rangle$  state while still exciting the $\left|  +  \right\rangle$  state. Thus, the degree of excitation of a given collective state is related to the degree of excitation of all other collective states. While the three-level system we are considering is more complicated in the details, this fundamental rule still holds. As such, under this set of conditions (i.e. semiclassical laser field, and no interaction between the atoms) it is not possible to block the excitation to state $\left|  {C_2}  \right\rangle$  while allowing for excitation of state $\left|  {C_1}  \right\rangle$. The result shown in Fig.~\ref{fig:fig4b} is merely a manifestation of this constraint. The subtle error that led us to the previous conclusion about the realizability of LSB was the approximation that the role of   $\left|  {G_2}  \right\rangle$ is negligible. This approximation was entirely logical in a general sense, but turns out, rather surprisingly, not to be valid.

   Of course, if the laser field is treated quantum mechanically, by considering it as a superposition of Fock states, the quantum state of the atoms and the photons are inherently entangled. As such, the state of the ensemble cannot be expressed as a product of the states of each atom. Under such a situation, it should in principle be possible to achieve the blockade effect. However, such a blockade works in a clean manner only when the numbers of photons are limited to a few. As discussed earlier, our objective is to achieve a blockade when the laser field has a mean photon number much larger than unity, i.e. the semi-classical limit. In this limit, the only way to achieve a blockade is to allow for interaction between the atoms. Here we describe a scheme where interactions between Rydberg excited levels are used to achieve the LSB effect.

\begin{figure}
        \centering
        \begin{subfigure}[b]{0.25\textwidth}
                \caption{}
                \includegraphics[width=\textwidth]{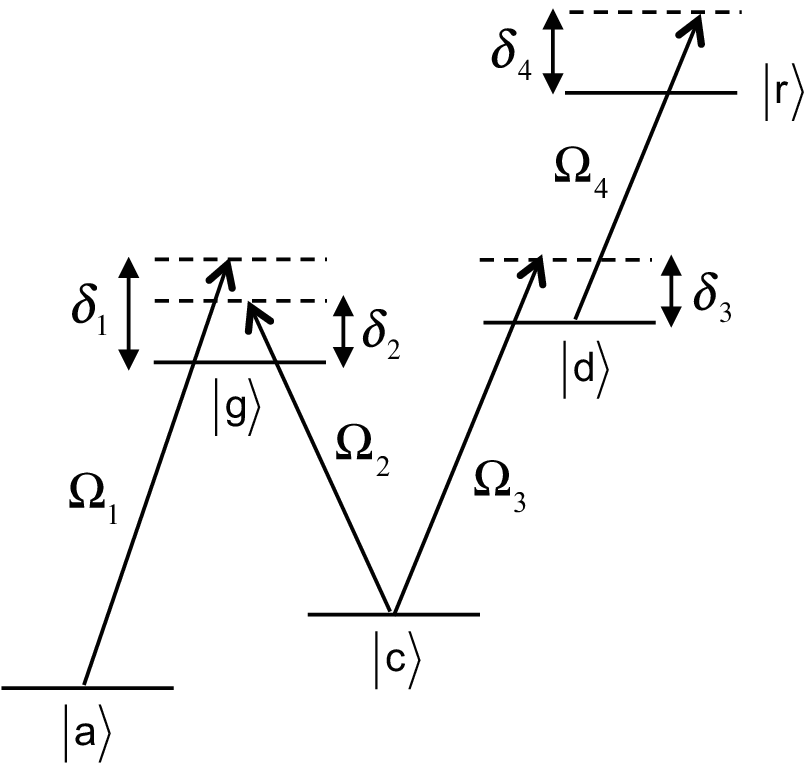}      
                \label{fig:fig5a}
        \end{subfigure}
        \begin{subfigure}[b]{0.4\textwidth}
                \caption{}
                \includegraphics[width=\textwidth]{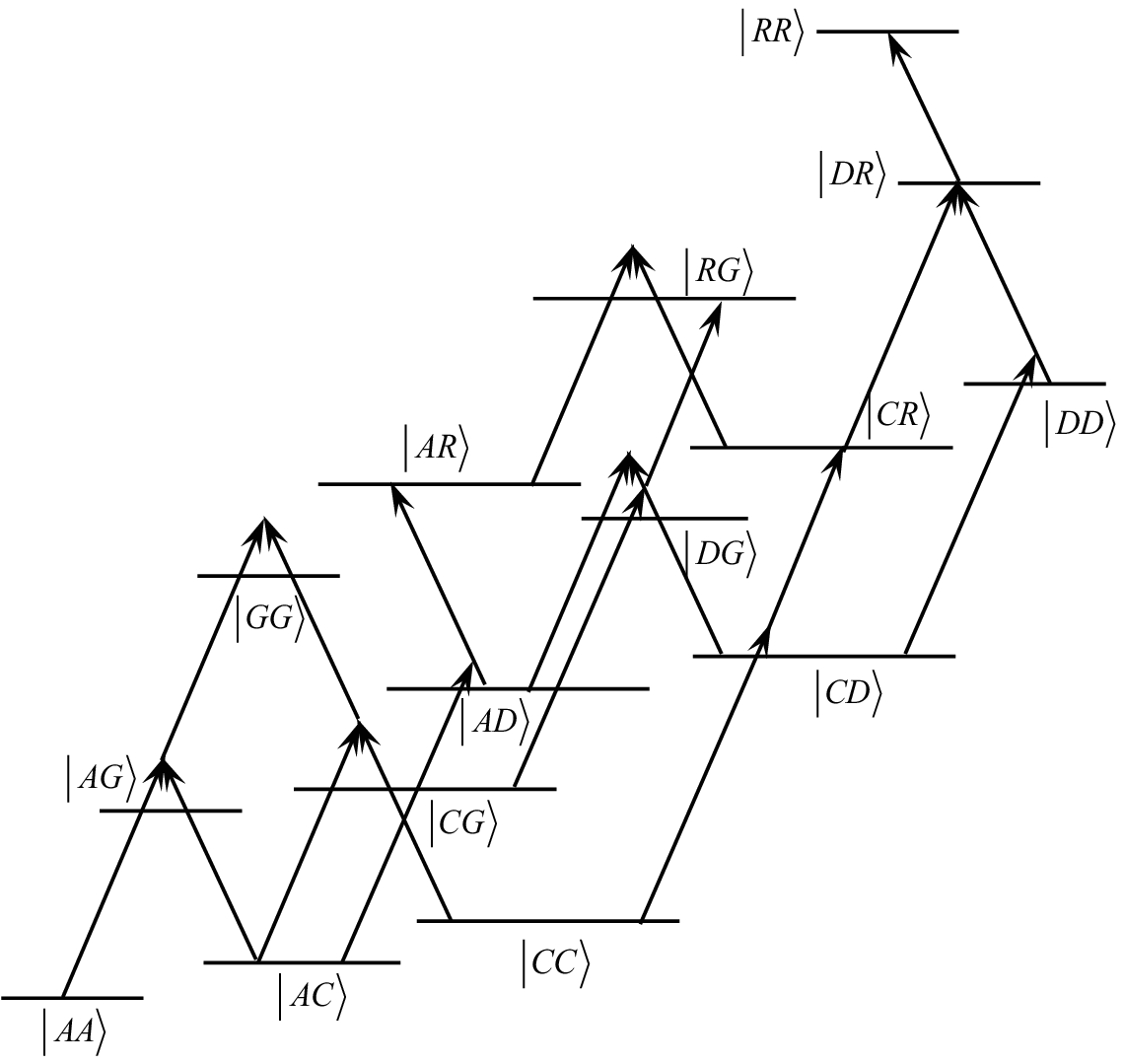}                
                \label{fig:fig5b}
        \end{subfigure}
\caption{(a) Modified $\Lambda$-system of a single atom. (b) Collective states of two atoms.}
\label{fig:fig5}
\end{figure}
   
\section{\label{sec:twoatom}Rydberg Assisted LSB of Two Atoms}
We modify the lambda scheme of a single atom by adding a Rydberg level $\left|  r  \right\rangle$ and an intermediate level $\left|  d  \right\rangle$, which is coupled to $\left|  r  \right\rangle$ and $\left|  c  \right\rangle$, but not to $\left|  a  \right\rangle$, as illustrated in Fig.~\ref{fig:fig5a}. We denote as $\hbar {\omega _j}$ the energy of the state $\left|  j  \right\rangle$, for $j=a$, $g$, $c$, $d$ and $r$.  The Rabi frequencies are denoted as $\Omega_1$, $\Omega_2$, $\Omega_3$ and $\Omega_4$ for the $a \to g$, $g \to c$, $c \to d$ and $d \to r$ transitions, respectively. For convenience, we also define the relevant detunings as ${\delta _1} = {\omega _1} - \left( {{\omega _b} - {\omega _a}} \right)$, ${\delta _2} = {\omega _2} - \left( {{\omega _b} - {\omega _c}} \right)$, ${\delta _3} = {\omega _3} - \left( {{\omega _d} - {\omega _c}} \right)$ and ${\delta _4} = {\omega _4} - \left( {{\omega _r} - {\omega _d}} \right)$. As before, the average detuning for the  $\Lambda$-transition is defined as $\delta  = {{\left( {{\delta _1} + {\delta _2}} \right)} \mathord{\left/
 {\vphantom {{\left( {{\delta _1} + {\delta _2}} \right)} 2}} \right.
 \kern-\nulldelimiterspace} 2}$, and the corresponding two photon detuning is defined as 
$\Delta  = {\delta _2} - {\delta _1}$. We also define as ${\delta _r} = {\delta _3} + {\delta _4}$ to be the two photon detuning for the ladder transition $c \to d \to r$. After making the usual dipole and rotating wave approximations and upon making the rotating wave transformation, the Hamiltonian in the basis of states $\left|  a  \right\rangle$, $\left|  g  \right\rangle$, $\left|  c  \right\rangle$, $\left|  d  \right\rangle$ and $\left|  r  \right\rangle$ can be expressed as
\begin{equation}
{H_{1R}} = \hbar \left[ {\begin{array}{*{20}{c}}
\Delta &{{{{\Omega _1}} \mathord{\left/
 {\vphantom {{{\Omega _1}} 2}} \right.
 \kern-\nulldelimiterspace} 2}}&0&0&0\\
{{{{\Omega _1}} \mathord{\left/
 {\vphantom {{{\Omega _1}} 2}} \right.
 \kern-\nulldelimiterspace} 2}}&{ - \delta  + {\Delta  \mathord{\left/
 {\vphantom {\Delta  2}} \right.
 \kern-\nulldelimiterspace} 2}}&{{{{\Omega _2}} \mathord{\left/
 {\vphantom {{{\Omega _2}} 2}} \right.
 \kern-\nulldelimiterspace} 2}}&0&0\\
0&{{{{\Omega _2}} \mathord{\left/
 {\vphantom {{{\Omega _2}} 2}} \right.
 \kern-\nulldelimiterspace} 2}}&0&{{{{\Omega _3}} \mathord{\left/
 {\vphantom {{{\Omega _3}} 2}} \right.
 \kern-\nulldelimiterspace} 2}}&0\\
0&0&{{{{\Omega _3}} \mathord{\left/
 {\vphantom {{{\Omega _3}} 2}} \right.
 \kern-\nulldelimiterspace} 2}}&{ - {\delta _3}}&{{{{\Omega _4}} \mathord{\left/
 {\vphantom {{{\Omega _4}} 2}} \right.
 \kern-\nulldelimiterspace} 2}}\\
0&0&0&{{{{\Omega _4}} \mathord{\left/
 {\vphantom {{{\Omega _4}} 2}} \right.
 \kern-\nulldelimiterspace} 2}}&{ - {\delta _r}}
\end{array}} \right].
\label{eq:eqn13}
\end{equation}
To illustrate the basic concept, we consider first the collective sates of only two atoms, with a distance $r_{12}$  which is assumed to be comparable to the characteristic distance scale of interatomic Rydberg interaction.
\begin{figure}
        \centering
        \begin{subfigure}[b]{0.25\textwidth}
                \caption{}
                \includegraphics[width=\textwidth]{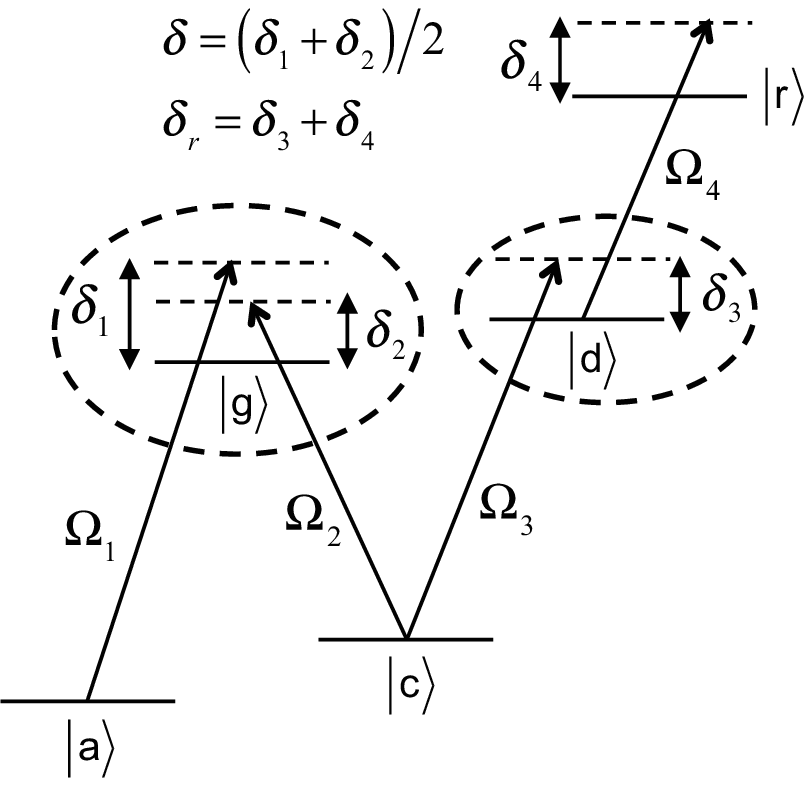}      
                \label{fig:fig6a}
        \end{subfigure}
        \qquad
        \begin{subfigure}[b]{0.175\textwidth}
                \caption{}
                \includegraphics[width=\textwidth]{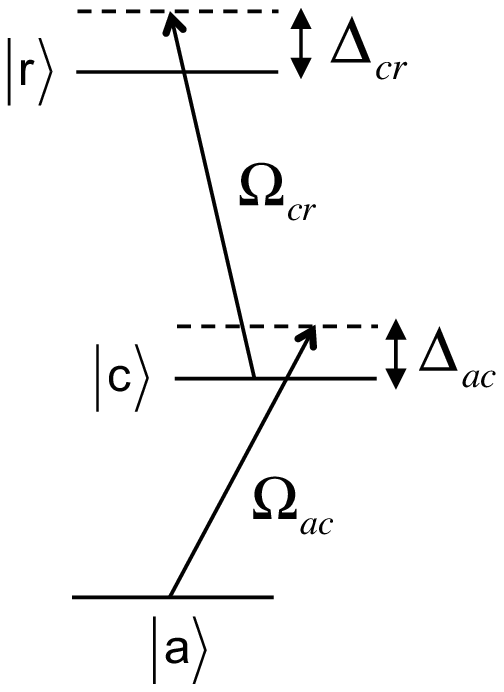}                
                \label{fig:fig6b}
        \end{subfigure}
\caption{(a) Single atom five-level scheme. (b) Simplified three-level scheme after adiabatically eliminating $\left| g \right\rangle$ and $\left| d \right\rangle $.}
\label{fig:fig6}
\end{figure}

\begin{figure*}
\includegraphics[width=0.7\textwidth]{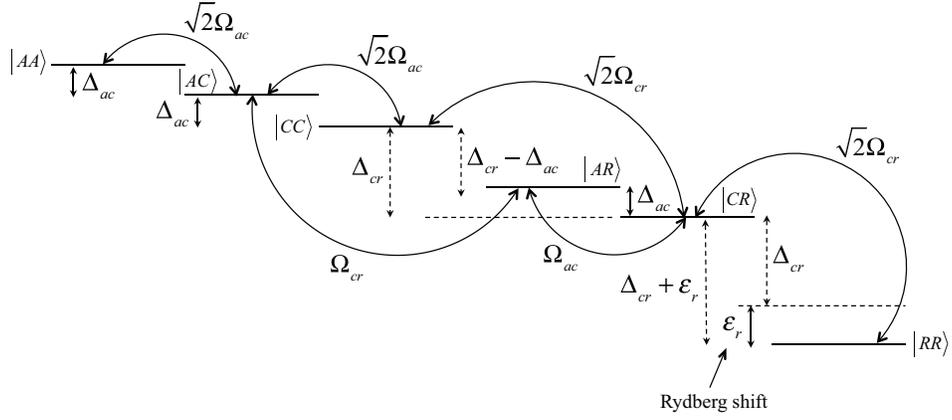}
\caption{The coupling rates and detunings of collective states of a simplified two-atom system.}
\label{fig:fig7}
\end{figure*}
   For simplicity, we consider first the symmetric collective states of two atoms, as illustrated in Fig.~\ref{fig:fig5b}, where we have adopted the compact notation that, for example, $\left| {AA} \right\rangle  = \left| {aa} \right\rangle $, $\left| {CC} \right\rangle  = \left| {cc} \right\rangle $, $\left| {AC} \right\rangle  = {{\left( {\left| {ac} \right\rangle  + \left| {ca} \right\rangle } \right)} \mathord{\left/ {\vphantom {{\left( {\left| {ac} \right\rangle  + \left| {ca} \right\rangle } \right)} {\sqrt 2 }}} \right. \kern-\nulldelimiterspace} {\sqrt 2 }}$  and so on. Since the Hamiltonian for the two atoms now contains the interaction between the two atoms, the general quantum state of the total system can no longer be written as a product between the quantum states of individual atoms. As such, it should now be possible to produce the LSB effect. Specifically, note that the dipole-dipole interaction between the atoms when they are both excited to the Rydberg state will shift the energy of the $\left| {RR} \right\rangle $  state compared to its value when the atoms are far apart. Since there is an asymmetry in the degree to which the $\left| r \right\rangle $  state is coupled to $\left| a \right\rangle $  and $\left| c \right\rangle $ , the shift in the energy of $\left| {RR} \right\rangle $  will affect differently the light shifts experienced by $\left| {AA} \right\rangle $, $\left| {AC} \right\rangle $  and $\left| {CC} \right\rangle $ . This is precisely what is needed for realizing LSB. In what follows, we derive analytically, under adiabatic elimination approximation, the parameters needed for realizing the optimal LSB condition. We then verify the results via exact numerical calculation. This is followed by a derivation of the condition needed for optimal LSB for an arbitrary value of  $N$, the number of atoms in the ensemble. 

   As can be seen from Fig.~\ref{fig:fig5b}, there are fifteen symmetric collective states for two atoms. In order to establish an approximate analytical result (which would then serve as a guide for choosing parameters for exact numerical calculation), we first simplify the picture by reducing the 5-level system for each atom (see Fig.~\ref{fig:fig6a}) to an effective 3-level system (see Fig.~\ref{fig:fig6b}) via eliminating adiabatically two of the intermediate states, $\left| g \right\rangle $   and $\left| d \right\rangle $  , that are highly detuned. Once this is done, the effective Hamiltonian for each atom, in the basis of $\left| a \right\rangle $  , $\left| c \right\rangle $   and $\left| r \right\rangle $  , can be expressed as       
\begin{equation}   
H{'_{1R}} = \hbar \left[ {\begin{array}{*{20}{c}}
{\Delta  + {\varepsilon _a}}&{{{{\Omega _{ac}}} \mathord{\left/
 {\vphantom {{{\Omega _{ac}}} 2}} \right.
 \kern-\nulldelimiterspace} 2}}&0\\
{{{{\Omega _{ac}}} \mathord{\left/
 {\vphantom {{{\Omega _{ac}}} 2}} \right.
 \kern-\nulldelimiterspace} 2}}&{{\varepsilon _c}}&{{{{\Omega _{cr}}} \mathord{\left/
 {\vphantom {{{\Omega _{cr}}} 2}} \right.
 \kern-\nulldelimiterspace} 2}}\\
0&{{{{\Omega _{cr}}} \mathord{\left/
 {\vphantom {{{\Omega _{cr}}} 2}} \right.
 \kern-\nulldelimiterspace} 2}}&{ - {\delta _r} + {\varepsilon _r}}
\end{array}} \right],
\label{eq:eqn14}
\end{equation}   
where ${\Omega _{ac}} = {{{\Omega _1}{\Omega _2}} \mathord{\left/
 {\vphantom {{{\Omega _1}{\Omega _2}} {2\delta }}} \right.
 \kern-\nulldelimiterspace} {2\delta }}$ is the Raman-Rabi frequency of transition $\left| a \right\rangle  \to \left| c \right\rangle $, and ${\Omega _{cr}} = {{{\Omega _3}{\Omega _4}} \mathord{\left/ {\vphantom {{{\Omega _3}{\Omega _4}} {2{\delta _3}}}} \right.
 \kern-\nulldelimiterspace} {2{\delta _3}}}$ is the two-photon Rabi frequency of transition $\left| c \right\rangle  \to \left| r \right\rangle $, while ${\varepsilon _a} = {{{\Omega _1}^2} \mathord{\left/ {\vphantom {{{\Omega _1}^2} {4\delta }}} \right. \kern-\nulldelimiterspace} {4\delta }}$, ${\varepsilon _c} = {{{\Omega _2}^2} \mathord{\left/
 {\vphantom {{{\Omega _2}^2} {4\delta }}} \right.
 \kern-\nulldelimiterspace} {4\delta }} + {{{\Omega _3}^2} \mathord{\left/
 {\vphantom {{{\Omega _3}^2} {4{\delta _3}}}} \right.
 \kern-\nulldelimiterspace} {4{\delta _3}}}$ and ${\varepsilon _r} = {{{\Omega _4}^2} \mathord{\left/ {\vphantom {{{\Omega _4}^2} {4{\delta _3}}}} \right.
 \kern-\nulldelimiterspace} {4{\delta _3}}}$ are the light shifts of states $\left| a \right\rangle $  , $\left| c \right\rangle $   and $\left| r \right\rangle $ respectively. If we define two new parameters ${\Delta _{ac}} = \Delta  + {\varepsilon _a} - {\varepsilon _c}$ and ${\Delta _{cr}} = {\delta _r} + {\varepsilon _c} - {\varepsilon _r}$, these become the effective, relevant detunings between the levels. Then we can rewrite the single atom Hamiltonian in the basis of
$\left| a \right\rangle$, $\left| c \right\rangle$ and $\left| r \right\rangle $ as 
\begin{equation}
H{'_{1R}} = \hbar \left[ {\begin{array}{*{20}{c}}
{{\Delta _{ac}}}&{{{{\Omega _{ac}}} \mathord{\left/
 {\vphantom {{{\Omega _{ac}}} 2}} \right.
 \kern-\nulldelimiterspace} 2}}&0\\
{{{{\Omega _{ac}}} \mathord{\left/
 {\vphantom {{{\Omega _{ac}}} 2}} \right.
 \kern-\nulldelimiterspace} 2}}&0&{{{{\Omega _{cr}}} \mathord{\left/
 {\vphantom {{{\Omega _{cr}}} 2}} \right.
 \kern-\nulldelimiterspace} 2}}\\
0&{{{{\Omega _{cr}}} \mathord{\left/
 {\vphantom {{{\Omega _{cr}}} 2}} \right.
 \kern-\nulldelimiterspace} 2}}&{ - {\Delta _{cr}}}
\end{array}} \right].
\label{eq:eqn15}
\end{equation}  

If the distance between the two atoms, $r_{12}$, is much larger than the scale of Rydberg interaction, the combined Hamiltonian in the basis of the nine product states ($\left| {{a_1}{a_2}} \right\rangle$, $\left| {{a_1}{c_2}} \right\rangle$, $\left| {{a_1}{r_2}} \right\rangle$, $\left| {{c_1}{a_2}} \right\rangle$, $\left| {{c_1}{c_2}} \right\rangle$, $\left| {{c_1}{r_2}} \right\rangle$, $\left| {{r_1}{a_2}} \right\rangle$, $\left| {{r_1}{c_2}} \right\rangle$, $\left| {{r_1}{r_2}} \right\rangle$) can be written as ${H_T} = {H_{1R}}^\prime  \otimes {I_2} + {I_1} \otimes {H_{2R}}^\prime$, and the 81 elements of  $H_T$ can be easily calculated in the same manner as used in deriving Eqn. (10). When transformed to the collective state picture, the asymmetric states become decoupled, just as before, and we are left with a six state system spanned by $\left| {AA} \right\rangle $, $\left| {AC} \right\rangle $, $\left| {CC} \right\rangle $, $\left| {AR} \right\rangle $, $\left| {CR} \right\rangle $ and $\left| {RR} \right\rangle $ (using the compact notation introduced in Fig.~\ref{fig:fig5b}), which are shown in Fig.~\ref{fig:fig7}, and the Hamiltonian can be expressed as 
\begin{widetext}
\begin{equation}
{H_T}^\prime  = \hbar \left[ {\begin{array}{*{20}{c}}
{2{\Delta _{ac}}}&{{\textstyle{{\sqrt 2 } \over 2}}{\Omega _{ac}}}&0&0&0&0\\
{{\textstyle{{\sqrt 2 } \over 2}}{\Omega _{ac}}}&{{\Delta _{ac}}}&{{\textstyle{{\sqrt 2 } \over 2}}{\Omega _{ac}}}&{{\textstyle{1 \over 2}}{\Omega _{cr}}}&0&0\\
0&{{\textstyle{{\sqrt 2 } \over 2}}{\Omega _{ac}}}&0&0&{{\textstyle{{\sqrt 2 } \over 2}}{\Omega _{cr}}}&0\\
0&{{\textstyle{1 \over 2}}{\Omega _{cr}}}&0&{{\Delta _{ac}} - {\Delta _{cr}}}&{{\textstyle{1 \over 2}}{\Omega _{ac}}}&0\\
0&0&{{\textstyle{{\sqrt 2 } \over 2}}{\Omega _{cr}}}&{{\textstyle{1 \over 2}}{\Omega _{ac}}}&{ - {\Delta _{cr}}}&{{\textstyle{{\sqrt 2 } \over 2}}{\Omega _{cr}}}\\
0&0&0&0&{{\textstyle{{\sqrt 2 } \over 2}}{\Omega _{cr}}}&{ - 2{\Delta _{cr}}}
\end{array}} \right].
\label{eq:eqn16}
\end{equation}
\end{widetext}  

\begin{figure*}
        \centering
        \begin{subfigure}[b]{0.35\textwidth}
                \caption{}
                \includegraphics[width=\textwidth]{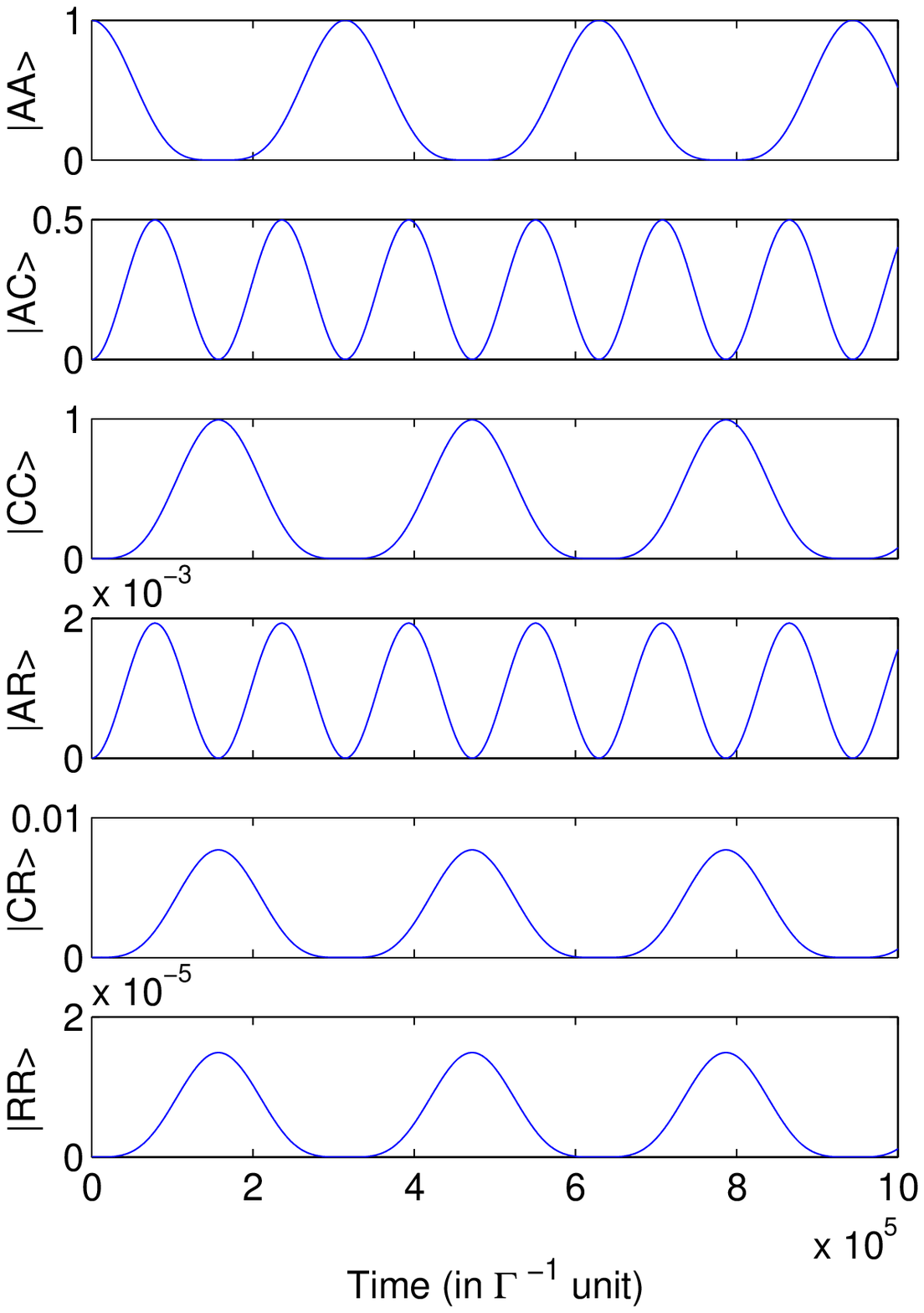}      
                \label{fig:fig8a}
        \end{subfigure}
        \qquad
        \begin{subfigure}[b]{0.35\textwidth}
                \caption{}
                \includegraphics[width=\textwidth]{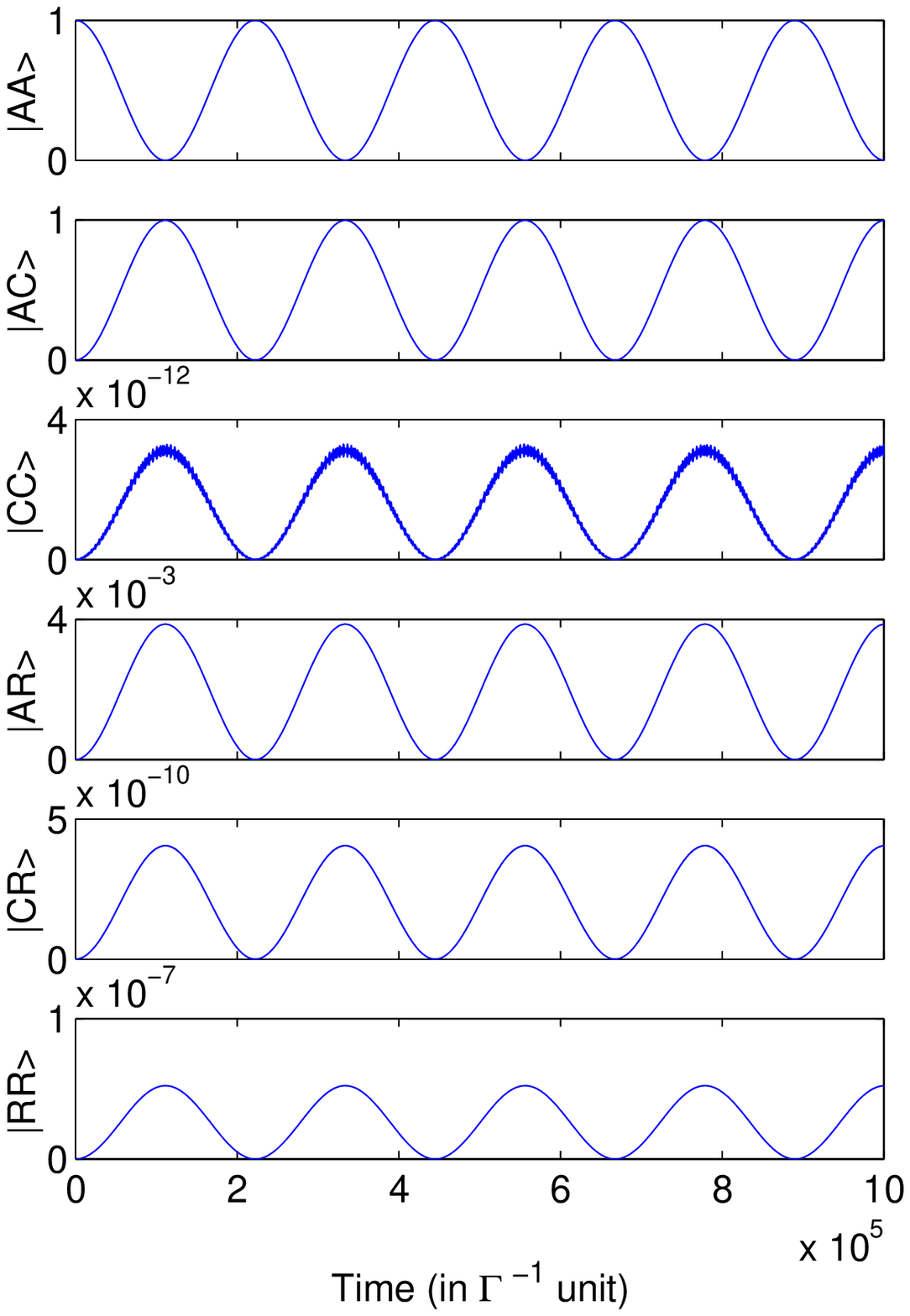}                
                \label{fig:fig8b}
        \end{subfigure}
\caption{Evolution of population using the simplified two-atom picture in Fig.~\ref{fig:fig6}. Figure (a) represents the case when the dipole-dipole interaction is not present $({V_r}=0 )$. Figure (b) represents the case when the dipole-dipole interaction is present $({V_r}=16 )$.}
\label{fig:fig8}
\end{figure*}
When the distance $r_{12}$   becomes comparable to the characteristic distance scale for interatomic Rydberg interaction, the Hamiltonian for the collective states, ${H_{TR}}^\prime$, is the same as  ${H_T}^\prime $ except for the last diagonal element. Specifically, $\left\langle {RR} \right|{H_{TR}}^\prime {\rm{ }}\left| {RR} \right\rangle  = \left\langle {RR} \right|{H_T}^\prime \left| {RR} \right\rangle  - {V_r} =  - 2{\Delta _{cr}} - {V_r}$, where $V_r$  represents the dipole-dipole interaction between two atoms. Thus, we can write
\begin{equation}
{H_{TR}}^\prime {\rm{ }} = {H_T}^\prime  - {V_R}\left| {RR} \right\rangle \left\langle {RR} \right|.
\label{eq:eqn17}
\end{equation}  
The various terms of  ${H_{TR}}^\prime $ are illustrated schematically in Fig.~\ref{fig:fig7}.

When we allow ${\Delta _{cr}} \gg {\Delta _{ac}}$ , ${\Omega _{ac}}$, ${\Omega _{cr}}$, the upper levels  $\left| {AR} \right\rangle $,  $\left| {CR} \right\rangle $ and $\left| {RR} \right\rangle $  can be adiabatically eliminated. The reduced Hamiltonian in the basis of $\left| {AA} \right\rangle $ , $\left| {AC} \right\rangle $  and  $\left| {CC} \right\rangle $ is 
\begin{widetext}
\begin{equation}
{\widetilde {{H_{TR}}}^\prime } \cong \hbar \left[ {\begin{array}{*{20}{c}}
{2{\Delta _{ac}}}&{\frac{{\sqrt 2 }}{2}{\Omega _{ac}}}&0\\
{\frac{{\sqrt 2 }}{2}{\Omega _{ac}}}&{{\Delta _{ac}} + \frac{{{\Omega _{cr}}}}{2} \cdot v}&{\frac{{\sqrt 2 }}{2}{\Omega _{ac}} + \frac{{\sqrt 2 }}{2}{\Omega _{cr}} \cdot \frac{{uv}}{{1 - 2vw}}}\\
0&{\frac{{\sqrt 2 }}{2}{\Omega _{ac}} + \frac{{\sqrt 2 }}{2}{\Omega _{cr}} \cdot \frac{{uv}}{{1 - 2vw}}}&{{\Omega _{cr}} \cdot \frac{v}{{1 - 2vw}}}
\end{array}} \right],
\label{eq:eqn18}
\end{equation} 
\end{widetext}
where, for simplicity, we have defined $u = {{{\Omega _{ac}}} \mathord{\left/
 {\vphantom {{{\Omega _{ac}}} {2{\Delta _{cr}}}}} \right.
 \kern-\nulldelimiterspace} {2{\Delta _{cr}}}}$, $v = {{{\Omega _{cr}}} \mathord{\left/
 {\vphantom {{{\Omega _{cr}}} {2{\Delta _{cr}}}}} \right.
 \kern-\nulldelimiterspace} {2{\Delta _{cr}}}}$, $w = {{{\Omega _{cr}}} \mathord{\left/
 {\vphantom {{{\Omega _{cr}}} {2\left( {2{\Delta _{cr}} + {V_r}} \right)}}} \right.
 \kern-\nulldelimiterspace} {2\left( {2{\Delta _{cr}} + {V_r}} \right)}}$, and we have assumed that ${\Omega _{cr}} \gg {\Omega _{ac}}$. In order to make the levels $\left| {AA} \right\rangle $   and  $\left| {AC} \right\rangle $  resonant, we enforce the condition that  ${\Delta _{ac}} = {{{\Omega _{cr}} \cdot v} \mathord{\left/
 {\vphantom {{{\Omega _{cr}} \cdot v} 2}} \right.
 \kern-\nulldelimiterspace} 2}$, which leads to $\Omega _{cr}^2 = 4{\Delta _{ac}}{\Delta _{cr}}$. When the energy levels are all reduced by  $2{\Delta _{ac}}$, Eqn.~(\ref{eq:eqn18}) becomes
\begin{widetext}
\begin{equation}
{\widetilde {{H_{TR}}}^\prime } = \hbar \left[ {\begin{array}{*{20}{c}}
0&{\frac{{\sqrt 2 }}{2}{\Omega _{ac}}}&0\\
{\frac{{\sqrt 2 }}{2}{\Omega _{ac}}}&0&{\frac{{\sqrt 2 }}{2}{\Omega _{ac}} + \frac{{\sqrt 2 }}{2}{\Omega _{cr}} \cdot \frac{{uv}}{{1 - 2vw}}}\\
0&{\frac{{\sqrt 2 }}{2}{\Omega _{ac}} + \frac{{\sqrt 2 }}{2}{\Omega _{cr}} \cdot \frac{{uv}}{{1 - 2vw}}}&{{\Delta _B}}
\end{array}} \right],
\label{eq:eqn19}
\end{equation} 
\end{widetext}
where ${\Delta _B} \equiv {{{\Omega _{cr}} \cdot 2vw} \mathord{\left/
 {\vphantom {{{\Omega _{cr}} \cdot 2vw} {\left( {1 - 2vw} \right)}}} \right.
 \kern-\nulldelimiterspace} {\left( {1 - 2vw} \right)}}$ is the blockade shift. When  ${\Delta _B}$ is much larger than the coupling between the states  $\left| {AC} \right\rangle$ and $\left| {CC} \right\rangle$, we are able to block the excitation to state  $\left| {CC} \right\rangle$ and achieve LSB. This can be achieved under the condition where ${V_r} + 2{\Delta _{cr}} \ll {{{\Omega _{cr}}} \mathord{\left/ {\vphantom {{{\Omega _{cr}}} {2{\Omega _{ac}}{\Delta _{cr}}}}} \right. \kern-\nulldelimiterspace} {2{\Omega _{ac}}{\Delta _{cr}}}}$. When these conditions are met, we achieve resonance between states  $\left| {AA} \right\rangle$ and $\left| {AC} \right\rangle$, blocking excitation to state $\left| {CC} \right\rangle$. 
 
   In order to verify the validity of this conclusion, we have simulated the evolution of the three-level system of two atoms (i.e. the system shown in Fig.~\ref{fig:fig6b}), using the 6 $\times$ 6  collective state Hamiltonian, ${H_{TR}}^\prime$ (Eqn.~(\ref{eq:eqn17})), which included the effect of Rydberg interaction, but without making use of the adiabatic elimination of states $\left| {AR} \right\rangle$, $\left| {CR} \right\rangle$  and $\left| {RR} \right\rangle$. The parameters we have used are ${\Omega _{ac}} = 0.00002$, ${\Omega _{cr}} = 1$, ${\Delta _{ac}} =  - 0.031129$ and ${\Delta _{cr}} =  - 8$ (in units of $\Gamma$), consistent with the requirement of achieving LSB. The result of this simulation is shown in Fig.~\ref{fig:fig8}. Fig.~\ref{fig:fig8a} represents the case when the Rydberg-interaction parameter, $V_r$ is set to zero. In this case, the maximum amplitude of $\left| {CC} \right\rangle$  reaches unity. When $V_r=16$, the maximum amplitude of  $\left| {CC} \right\rangle$ is nearly zero, and the system oscillates between $\left| {AA} \right\rangle$  and $\left| {AC} \right\rangle$, as shown in Fig.~\ref{fig:fig8b}. It should also be noted that under this blockade condition, the oscillation frequency between levels   $\left| {AA} \right\rangle$  and $\left| {AC} \right\rangle$  is increased by  $\sqrt 2$. The upper levels $\left| {AR} \right\rangle$, $\left| {CR} \right\rangle$  and $\left| {RR} \right\rangle$ are minimally excited regardless of whether interaction is present or not. This justifies the adiabatic elimination of these states employed in deriving the 3$\times$3  reduced Hamiltonian for the collective states, shown in Eqn.~(\ref{eq:eqn18}).

\begin{figure*}
\centering
\includegraphics[width=0.7\textwidth]{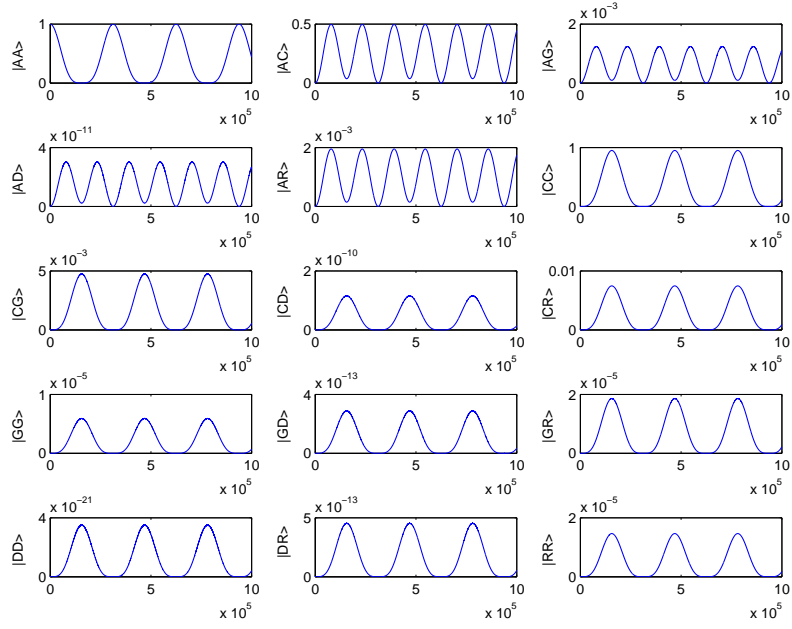}
\caption{Evolution of population using the full two-atom picture in Fig.~\ref{fig:fig6b} when the dipole-dipole interaction is not present  $({V_r}=0)$.}
\label{fig:fig9}
\end{figure*}

\begin{figure*}
\includegraphics[width=0.7\textwidth]{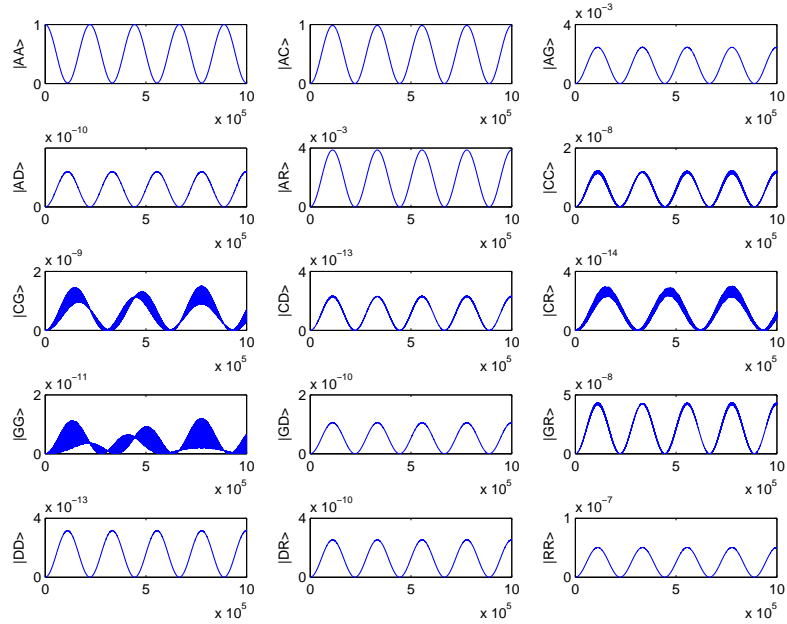}
\caption{Evolution of population using the full two-atom picture in Fig.~\ref{fig:fig6b} when the dipole-dipole interaction is present  $({V_r}=16)$.}
\label{fig:fig10}
\end{figure*}
 
   The parameters used in the evolution of the simplified two-atom Hamiltonian can be used to extract the values of parameters necessary for the exact two-atom 15-level system shown in Fig.~\ref{fig:fig5b}. We choose the parameters as follows: ${\Omega _1} = 0.0004$, ${\Omega _2} = 0.8$, $\delta  =  - 8$, $\Delta  =  - {\rm{0.0199}}$, ${\Omega _3} = 20$, ${\Omega _4} = 320$, ${\delta _3} =  - 3200$, ${\delta _4} = 3200$. Notice that here we make the choice that $\Delta  \simeq {{\left( {{\Omega _2}^2 - {\Omega _1}^2} \right)} \mathord{\left/
 {\vphantom {{\left( {{\Omega _2}^2 - {\Omega _1}^2} \right)} {4\delta }}} \right.
 \kern-\nulldelimiterspace} {4\delta }}$  in order to produce full Rabi oscillations between  $\left| {AA} \right\rangle$  and $\left| {AC} \right\rangle$. The results of the plots with and without the Rydberg interaction are shown in Fig.~\ref{fig:fig9}. Despite the fact that 15 levels are present, only the levels  $\left| {AA} \right\rangle$, $\left| {AC} \right\rangle$ and $\left| {CC} \right\rangle$  are populated while the excitations to the other states remain under 1\%. As was the case with the simplified Hamiltonian, the presence of the Rydberg interaction (${V_r} = 16 $) suppresses the excitation to level $\left| {CC} \right\rangle$ so that an effective two-level system is generated, as illustrated in Fig.~\ref{fig:fig10}. 

\section{\label{sec:Natom}Rydberg Assisted LSB in $N$-atom ensembles}
This process can be generalized for $N$ atoms. Referring back to Fig.~\ref{fig:fig6}, we recall first that adiabatic elimination of states  $\left| g \right\rangle$ and $\left| d \right\rangle$  reduces the system to three levels (Fig.~\ref{fig:fig6b}). The first six collective states involving these single atom states, for $N$-atoms, are as follows
\begin{equation}
\begin{array}{l}
\left| A \right\rangle  \equiv \left| {{a_1},{a_2}, \cdot  \cdot ,{a_N}} \right\rangle ,\\
\left| {{C_1}} \right\rangle  \equiv \frac{1}{{\sqrt N }}\sum\limits_{j = 1}^N {\left| {{a_1},{a_2}, \cdot  \cdot ,{c_j}, \cdot  \cdot ,{a_N}} \right\rangle } ,\\
\left| {{C_2}} \right\rangle  \equiv \frac{1}{{\sqrt {{}^N{C_2}} }}\sum\limits_{j,k(j \ne k)}^{{}^N{C_2}} {\left| {{a_1},{a_2}, \cdot  \cdot ,{c_j}, \cdot  \cdot ,{c_k}, \cdot  \cdot ,{a_N}} \right\rangle } ,\\
\left| {{R_1}} \right\rangle  \equiv \frac{1}{{\sqrt N }}\sum\limits_{j = 1}^N {\left| {{a_1},{a_2}, \cdot  \cdot ,{r_j}, \cdot  \cdot ,{a_N}} \right\rangle } ,\\
\left| {{R_{1,1}}} \right\rangle  \equiv \frac{1}{{\sqrt {2{}^N{C_2}} }}\sum\limits_{j,k(j \ne k)}^{2{}^N{C_2}} {\left| {{a_1},{a_2}, \cdot  \cdot ,{r_j}, \cdot  \cdot ,{c_k}, \cdot  \cdot ,{a_N}} \right\rangle } ,\\
\left| {{R_2}} \right\rangle  \equiv \frac{1}{{\sqrt {{}^N{C_2}} }}\sum\limits_{j,k(j \ne k)}^{{}^N{C_2}} {\left| {{a_1},{a_2}, \cdot  \cdot ,{r_j}, \cdot  \cdot ,{r_k}, \cdot  \cdot ,{a_N}} \right\rangle } .
\end{array}
\label{eq:eqn20}
\end{equation} 

Of course, there are many more collective states. However, our goal is to find the condition where the system oscillates between  $\left| A \right\rangle$ and $\left| {C_1} \right\rangle$, with negligible excitation to the remaining collective states. If we can show that the excitation to states  $\left| {C_2} \right\rangle$,  $\left| {R_1} \right\rangle$, $\left| {R_{1,1}} \right\rangle$  and  $\left| {R_2} \right\rangle$ are negligible, then it follows that the excitation to all other higher energy collective states is also negligible. Thus, it is justified to limit our consideration to only these six states.

With the single atom Hamiltonian in the basis of  $\left| a \right\rangle$, $\left| c \right\rangle$  and $\left| r \right\rangle$  shown in Eqn.~(\ref{eq:eqn15}), the Hamiltonian formed with states  $\left| A \right\rangle$, $\left| {C_1} \right\rangle$,  $\left| {C_2} \right\rangle$,  $\left| {R_1} \right\rangle$, $\left| {R_{1,1}} \right\rangle$  and  $\left| {R_2} \right\rangle$  can be written as
\begin{widetext}
\begin{equation}
{H_{NR}}^\prime {\rm{ }} = \hbar \left[ {\begin{array}{*{20}{c}}
{2{\Delta _{ac}}}&{\frac{{\sqrt N }}{2}{\Omega _{ac}}}&0&0&0&0\\
{\frac{{\sqrt N }}{2}{\Omega _{ac}}}&{{\Delta _{ac}}}&{\frac{{\sqrt {2\left( {N - 1} \right)} }}{2}{\Omega _{ac}}}&{\frac{{{\Omega _{cr}}}}{2}}&0&0\\
0&{\frac{{\sqrt {2\left( {N - 1} \right)} }}{2}{\Omega _{ac}}}&0&0&{\frac{{\sqrt 2 }}{2}{\Omega _{cr}}}&0\\
0&{\frac{{{\Omega _{cr}}}}{2}}&0&{{\Delta _{ac}} - {\Delta _{cr}}}&{\frac{{\sqrt {N - 1} }}{2}{\Omega _{ac}}}&0\\
0&0&{\frac{{\sqrt 2 }}{2}{\Omega _{cr}}}&{\frac{{\sqrt {N - 1} }}{2}{\Omega _{ac}}}&{ - {\Delta _{cr}}}&{\frac{{\sqrt 2 }}{2}{\Omega _{cr}}}\\
0&0&0&0&{\frac{{\sqrt 2 }}{2}{\Omega _{cr}}}&{ - 2{\Delta _{cr}} - {V_r}}
\end{array}} \right].
\label{eq:eqn21}
\end{equation} 
\end{widetext}
Under the condition that ${\Delta _{cr}} \gg {\Delta _{ac}}$,$\sqrt N {\Omega _{ac}}$, ${\Omega _{cr}}$, for large $N$, this reduces to
\begin{widetext}
\begin{equation}
\widetilde {{H_{NR}}^\prime {\rm{ }}} = \hbar \left[ {\begin{array}{*{20}{c}}
0&{\frac{{\sqrt 2 }}{2}{\Omega _{ac}}}&0\\
{\frac{{\sqrt 2 }}{2}{\Omega _{ac}}}&0&{\frac{{\sqrt 2 }}{2}{\Omega _{ac}} + \frac{{{\Omega _{cr}}}}{2} \cdot \frac{{uv\sqrt {2\left( {N - 1} \right)} }}{{1 - \left( {N - 1} \right){u^2} - 2vw}}}\\
0&{\frac{{\sqrt 2 }}{2}{\Omega _{ac}} + \frac{{{\Omega _{cr}}}}{2} \cdot \frac{{uv\sqrt {2\left( {N - 1} \right)} }}{{1 - \left( {N - 1} \right){u^2} - 2vw}}}&{{\Delta _B}}
\end{array}} \right]
\label{eq:eqn22}
\end{equation} 
\end{widetext}
in the basis of $\left| A \right\rangle$ and $\left| {C_1} \right\rangle$ and $\left| {C_2} \right\rangle$, where the first two levels were made resonant by choosing ${\Delta _{ac}} = {{\left( {{{{\Omega _{cr}}} \mathord{\left/
 {\vphantom {{{\Omega _{cr}}} 2}} \right.
 \kern-\nulldelimiterspace} 2}} \right) \cdot v\left( {1 - 2vw} \right)} \mathord{\left/
 {\vphantom {{\left( {{{{\Omega _{cr}}} \mathord{\left/
 {\vphantom {{{\Omega _{cr}}} 2}} \right.
 \kern-\nulldelimiterspace} 2}} \right) \cdot v\left( {1 - 2vw} \right)} {\left( {1 - \left( {N - 1} \right){u^2} - 2vw} \right)}}} \right.
 \kern-\nulldelimiterspace} {\left( {1 - \left( {N - 1} \right){u^2} - 2vw} \right)}}$. The blockade shift is now ${\Delta _B} \equiv {{{\Omega _{cr}} \cdot 2vw} \mathord{\left/
 {\vphantom {{{\Omega _{cr}} \cdot 2vw} {\left( {1 - \left( {N - 1} \right){u^2} - 2vw} \right)}}} \right.
 \kern-\nulldelimiterspace} {\left( {1 - \left( {N - 1} \right){u^2} - 2vw} \right)}}$. Note that when $N = 2$, the Hamiltonian, the detuning, and the blockade shift are equivalent to the calculations made earlier for the two-atom case. The conditions necessary to block the excitation to state $\left| {C_2} \right\rangle$ are ${\Omega _{cr}} \gg \sqrt N {\Omega _{ac}}$ and $w \gg \sqrt N u$, which again occur when  ${V_r} \to  - 2{\Delta _{cr}}$, just as in the case of  $N = 2$. 

\begin{figure}
\includegraphics[trim=1.5cm 0cm 1cm 0cm, clip, width=8cm]{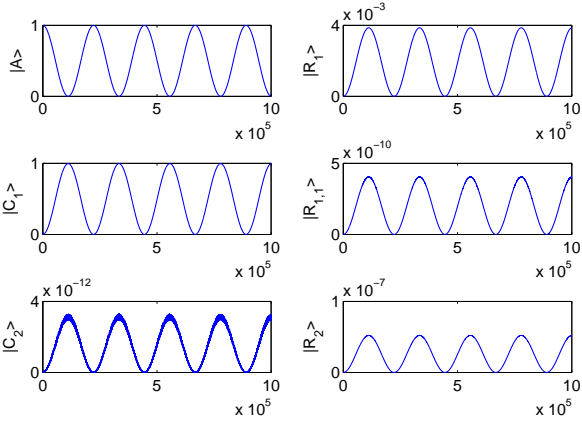}
\caption{Evolution of population in the six lowest energy states of Hamiltonian in Eqn.~(\ref{eq:eqn20}) for $N = 1000$, with the same conditions as Fig.~\ref{fig:fig8} except $\Omega_{ac}$ here is $\sqrt N$  smaller, and the dipole-dipole interaction ${V_r}=16\Gamma$.}
\label{fig:fig11}
\end{figure}

Fig.~\ref{fig:fig11} shows the populations of the six collective states of Eqn.~(\ref{eq:eqn20}) under the LSB conditions found for 1000 atoms. The parameters are  ${\Omega _{ac}} = {{0.00002} \mathord{\left/
 {\vphantom {{0.00002} {\sqrt {1000} }}} \right.
 \kern-\nulldelimiterspace} {\sqrt {1000} }}$,  ${\Omega _{cr}} = 1$, ${\Delta _{ac}} =  - 0.031129$,  ${\Delta _{cr}} =  - 8$ and ${V_r} = 16$  (in units of  $\Gamma$). As can be seen, states  $\left| A \right\rangle$ and $\left| {C_1} \right\rangle$  are resonant, and population in state  $\left| {C_2} \right\rangle$ is very small. With so little excitation into $\left| {C_2} \right\rangle$ , the Rydberg assisted LSB guarantees the suppression of the higher excitations, thereby validating the use of a truncated Hamiltonian in Eqn.~(\ref{eq:eqn20}).

\begin{figure}
\includegraphics[trim=1.5cm 0cm 1cm 0cm, clip, width=8cm]{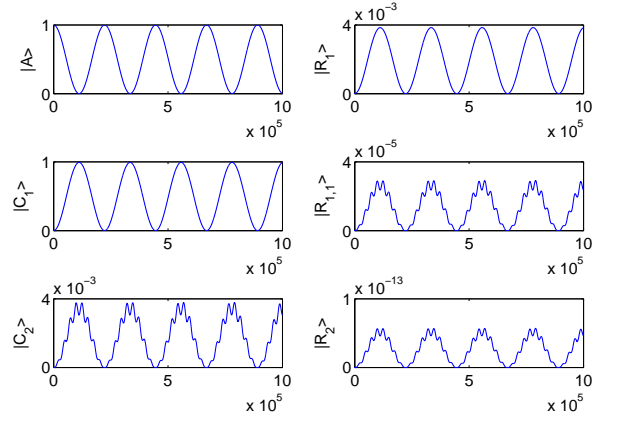}
\caption{Evolution of population in the six lowest energy states of Hamiltonian in Eqn.~(\ref{eq:eqn20}), with the same conditions as Fig.~\ref{fig:fig11} except the dipole-dipole interaction ${V_r}=16000\Gamma$.}
\label{fig:fig12}
\end{figure}

So far, we have shown that the Rydberg assisted LSB works for ${V_r} = 16\Gamma$, where $\Gamma$ is the decay rate of the state  $\left| g \right\rangle$. Consider, for example, the specific case of $^{87}$Rb atoms. In this case, $\Gamma  \simeq 6$MHz, so that ${V_r} \simeq 96$MHz, which corresponds to an interatomic distance of $\sim 10\mu m$. We envision a scenario where the collective ensemble would be confined to a sphere with a diameter $\sim 10\mu m$, realizable, for example, by loading atoms from a MOT into a FORT (far-off resonance trap), containing about  $10^3$ atoms. For some pair of atoms, the interatomic distance would be smaller than  $10\mu m$. It is well known that $V_r$  scales approximately as $r^{-3}$, where $r$ is the interatomic distance between a pair of atoms for $r < 10\mu m$~\cite{RevModPhys.82.2313}. Thus, for $r = 1\mu m$,  ${V_r} \simeq 16000\Gamma \simeq 96 \times {10^3}{\rm{MHz}}$. We show in Fig.~\ref{fig:fig12} that the Rydberg assisted LSB works for this value of $V_r$ for  $N = 1000$ atoms.

\section{\label{sec:conclusion}Conclusion}
   The light shift imbalance induced blockade in an atomic ensemble had been studied previously, in which the difference in the light shifts produced in collective state energy levels leads to a condition where the system remains confined to a superposition of the ground and the first excited states. The significance of this result for quantum computing was discussed in Ref~\cite{Steane:92}. Upon further investigation into the nature of collective states, we found that the light shift imbalance alone is not enough to produce a blockade. By introducing Rydberg interaction, and using the technique of adiabatic elimination, we are able to establish the conditions under which the blockade can be achieved. Numerical simulations confirm the validity of this result. 
      
The ensemble-based qubits realized in this manner can be used to implement a controlled-NOT (CNOT) gate, which is a universal gate for quantum computing, using a variation of the Pellizzari scheme~\cite{PhysRevLett.75.3788}. The details of the process for realizing a CNOT gate in this way, using $^{87}$Rb atoms are essentially the same as what was presented in Ref~\cite{PhysRevA.75.022323}. Many such gates can be linked to one another, via nearest neighbor quantum coupling, to realize an elementary quantum computer (EQC). The size of an EQC, contained inside a single vacuum chamber, is likely to be limited to a number of the order of ten. However, as shown in Ref~\cite{PhysRevA.75.022323}, many such EQCs can be linked via optical fiber, using photons to transport quantum information, thus making this approach scalable. Of course, it is also possible to realize a CNOT gate between single atoms, caught in FORTs, by making use of Rydberg interactions~\cite{SaffmanNatPhys}. However, it is very difficult to load a single atom consistently in a FORT. In contrast, the approach proposed here is relatively insensitive to the actual number of atoms held in the FORT. Thus, this approach may prove to be a more viable alternative for scalable quantum computing using neutral atoms.              

\begin{acknowledgments}
This work has been supported by the NSF IGERT program under grant number `DGE-0801685' and AFOSR grant number `FA9550-09-1-0652'.
\end{acknowledgments}

\nocite{*}
\bibliography{LSIIB}

\providecommand{\noopsort}[1]{}\providecommand{\singleletter}[1]{#1}%
\begin{thebibliography}{21}%
\makeatletter
\providecommand \@ifxundefined [1]{%
 \@ifx{#1\undefined}
}%
\providecommand \@ifnum [1]{%
 \ifnum #1\expandafter \@firstoftwo
 \else \expandafter \@secondoftwo
 \fi
}%
\providecommand \@ifx [1]{%
 \ifx #1\expandafter \@firstoftwo
 \else \expandafter \@secondoftwo
 \fi
}%
\providecommand \natexlab [1]{#1}%
\providecommand \enquote  [1]{``#1''}%
\providecommand \bibnamefont  [1]{#1}%
\providecommand \bibfnamefont [1]{#1}%
\providecommand \citenamefont [1]{#1}%
\providecommand \href@noop [0]{\@secondoftwo}%
\providecommand \href [0]{\begingroup \@sanitize@url \@href}%
\providecommand \@href[1]{\@@startlink{#1}\@@href}%
\providecommand \@@href[1]{\endgroup#1\@@endlink}%
\providecommand \@sanitize@url [0]{\catcode `\\12\catcode `\$12\catcode
  `\&12\catcode `\#12\catcode `\^12\catcode `\_12\catcode `\%12\relax}%
\providecommand \@@startlink[1]{}%
\providecommand \@@endlink[0]{}%
\providecommand \url  [0]{\begingroup\@sanitize@url \@url }%
\providecommand \@url [1]{\endgroup\@href {#1}{\urlprefix }}%
\providecommand \urlprefix  [0]{URL }%
\providecommand \Eprint [0]{\href }%
\providecommand \doibase [0]{http://dx.doi.org/}%
\providecommand \selectlanguage [0]{\@gobble}%
\providecommand \bibinfo  [0]{\@secondoftwo}%
\providecommand \bibfield  [0]{\@secondoftwo}%
\providecommand \translation [1]{[#1]}%
\providecommand \BibitemOpen [0]{}%
\providecommand \bibitemStop [0]{}%
\providecommand \bibitemNoStop [0]{.\EOS\space}%
\providecommand \EOS [0]{\spacefactor3000\relax}%
\providecommand \BibitemShut  [1]{\csname bibitem#1\endcsname}%
\let\auto@bib@innerbib\@empty
\bibitem [{\citenamefont {Jaksch}\ \emph {et~al.}(2000)\citenamefont {Jaksch},
  \citenamefont {Cirac}, \citenamefont {Zoller}, \citenamefont {Rolston},
  \citenamefont {C\^ot\'e},\ and\ \citenamefont {Lukin}}]{PhysRevLett.85.2208}%
  \BibitemOpen
  \bibfield  {author} {\bibinfo {author} {\bibfnamefont {D.}~\bibnamefont
  {Jaksch}}, \bibinfo {author} {\bibfnamefont {J.~I.}\ \bibnamefont {Cirac}},
  \bibinfo {author} {\bibfnamefont {P.}~\bibnamefont {Zoller}}, \bibinfo
  {author} {\bibfnamefont {S.~L.}\ \bibnamefont {Rolston}}, \bibinfo {author}
  {\bibfnamefont {R.}~\bibnamefont {C\^ot\'e}}, \ and\ \bibinfo {author}
  {\bibfnamefont {M.~D.}\ \bibnamefont {Lukin}},\ }\href {\doibase
  10.1103/PhysRevLett.85.2208} {\bibfield  {journal} {\bibinfo  {journal}
  {Phys. Rev. Lett.}\ }\textbf {\bibinfo {volume} {85}},\ \bibinfo {pages}
  {2208} (\bibinfo {year} {2000})}\BibitemShut {NoStop}%
\bibitem [{\citenamefont {Lukin}\ \emph {et~al.}(2001)\citenamefont {Lukin},
  \citenamefont {Fleischhauer}, \citenamefont {Cote}, \citenamefont {Duan},
  \citenamefont {Jaksch}, \citenamefont {Cirac},\ and\ \citenamefont
  {Zoller}}]{PhysRevLett.87.037901}%
  \BibitemOpen
  \bibfield  {author} {\bibinfo {author} {\bibfnamefont {M.~D.}\ \bibnamefont
  {Lukin}}, \bibinfo {author} {\bibfnamefont {M.}~\bibnamefont {Fleischhauer}},
  \bibinfo {author} {\bibfnamefont {R.}~\bibnamefont {Cote}}, \bibinfo {author}
  {\bibfnamefont {L.~M.}\ \bibnamefont {Duan}}, \bibinfo {author}
  {\bibfnamefont {D.}~\bibnamefont {Jaksch}}, \bibinfo {author} {\bibfnamefont
  {J.~I.}\ \bibnamefont {Cirac}}, \ and\ \bibinfo {author} {\bibfnamefont
  {P.}~\bibnamefont {Zoller}},\ }\href {\doibase 10.1103/PhysRevLett.87.037901}
  {\bibfield  {journal} {\bibinfo  {journal} {Phys. Rev. Lett.}\ }\textbf
  {\bibinfo {volume} {87}},\ \bibinfo {pages} {037901} (\bibinfo {year}
  {2001})}\BibitemShut {NoStop}%
\bibitem [{\citenamefont {Pritchard}\ \emph {et~al.}(2010)\citenamefont
  {Pritchard}, \citenamefont {Maxwell}, \citenamefont {Gauguet}, \citenamefont
  {Weatherill}, \citenamefont {Jones},\ and\ \citenamefont
  {Adams}}]{PhysRevLett.105.193603}%
  \BibitemOpen
  \bibfield  {author} {\bibinfo {author} {\bibfnamefont {J.~D.}\ \bibnamefont
  {Pritchard}}, \bibinfo {author} {\bibfnamefont {D.}~\bibnamefont {Maxwell}},
  \bibinfo {author} {\bibfnamefont {A.}~\bibnamefont {Gauguet}}, \bibinfo
  {author} {\bibfnamefont {K.~J.}\ \bibnamefont {Weatherill}}, \bibinfo
  {author} {\bibfnamefont {M.~P.~A.}\ \bibnamefont {Jones}}, \ and\ \bibinfo
  {author} {\bibfnamefont {C.~S.}\ \bibnamefont {Adams}},\ }\href {\doibase
  10.1103/PhysRevLett.105.193603} {\bibfield  {journal} {\bibinfo  {journal}
  {Phys. Rev. Lett.}\ }\textbf {\bibinfo {volume} {105}},\ \bibinfo {pages}
  {193603} (\bibinfo {year} {2010})}\BibitemShut {NoStop}%
\bibitem [{\citenamefont {Heidemann}\ \emph {et~al.}(2007)\citenamefont
  {Heidemann}, \citenamefont {Raitzsch}, \citenamefont {Bendkowsky},
  \citenamefont {Butscher}, \citenamefont {L\"ow}, \citenamefont {Santos},\
  and\ \citenamefont {Pfau}}]{PhysRevLett.99.163601}%
  \BibitemOpen
  \bibfield  {author} {\bibinfo {author} {\bibfnamefont {R.}~\bibnamefont
  {Heidemann}}, \bibinfo {author} {\bibfnamefont {U.}~\bibnamefont {Raitzsch}},
  \bibinfo {author} {\bibfnamefont {V.}~\bibnamefont {Bendkowsky}}, \bibinfo
  {author} {\bibfnamefont {B.}~\bibnamefont {Butscher}}, \bibinfo {author}
  {\bibfnamefont {R.}~\bibnamefont {L\"ow}}, \bibinfo {author} {\bibfnamefont
  {L.}~\bibnamefont {Santos}}, \ and\ \bibinfo {author} {\bibfnamefont
  {T.}~\bibnamefont {Pfau}},\ }\href {\doibase 10.1103/PhysRevLett.99.163601}
  {\bibfield  {journal} {\bibinfo  {journal} {Phys. Rev. Lett.}\ }\textbf
  {\bibinfo {volume} {99}},\ \bibinfo {pages} {163601} (\bibinfo {year}
  {2007})}\BibitemShut {NoStop}%
\bibitem [{\citenamefont {Dudin}\ and\ \citenamefont
  {Kuzmich}(2012)}]{Dudin18052012}%
  \BibitemOpen
  \bibfield  {author} {\bibinfo {author} {\bibfnamefont {Y.~O.}\ \bibnamefont
  {Dudin}}\ and\ \bibinfo {author} {\bibfnamefont {A.}~\bibnamefont
  {Kuzmich}},\ }\href {\doibase 10.1126/science.1217901} {\bibfield  {journal}
  {\bibinfo  {journal} {Science}\ }\textbf {\bibinfo {volume} {336}},\ \bibinfo
  {pages} {887} (\bibinfo {year} {2012})}\BibitemShut {NoStop}%
\bibitem [{\citenamefont {M\o{}ller}\ \emph {et~al.}(2008)\citenamefont
  {M\o{}ller}, \citenamefont {Madsen},\ and\ \citenamefont
  {M\o{}lmer}}]{PhysRevLett.100.170504}%
  \BibitemOpen
  \bibfield  {author} {\bibinfo {author} {\bibfnamefont {D.}~\bibnamefont
  {M\o{}ller}}, \bibinfo {author} {\bibfnamefont {L.~B.}\ \bibnamefont
  {Madsen}}, \ and\ \bibinfo {author} {\bibfnamefont {K.}~\bibnamefont
  {M\o{}lmer}},\ }\href {\doibase 10.1103/PhysRevLett.100.170504} {\bibfield
  {journal} {\bibinfo  {journal} {Phys. Rev. Lett.}\ }\textbf {\bibinfo
  {volume} {100}},\ \bibinfo {pages} {170504} (\bibinfo {year}
  {2008})}\BibitemShut {NoStop}%
\bibitem [{\citenamefont {Vogt}\ \emph {et~al.}(2006)\citenamefont {Vogt},
  \citenamefont {Viteau}, \citenamefont {Zhao}, \citenamefont {Chotia},
  \citenamefont {Comparat},\ and\ \citenamefont
  {Pillet}}]{PhysRevLett.97.083003}%
  \BibitemOpen
  \bibfield  {author} {\bibinfo {author} {\bibfnamefont {T.}~\bibnamefont
  {Vogt}}, \bibinfo {author} {\bibfnamefont {M.}~\bibnamefont {Viteau}},
  \bibinfo {author} {\bibfnamefont {J.}~\bibnamefont {Zhao}}, \bibinfo {author}
  {\bibfnamefont {A.}~\bibnamefont {Chotia}}, \bibinfo {author} {\bibfnamefont
  {D.}~\bibnamefont {Comparat}}, \ and\ \bibinfo {author} {\bibfnamefont
  {P.}~\bibnamefont {Pillet}},\ }\href {\doibase 10.1103/PhysRevLett.97.083003}
  {\bibfield  {journal} {\bibinfo  {journal} {Phys. Rev. Lett.}\ }\textbf
  {\bibinfo {volume} {97}},\ \bibinfo {pages} {083003} (\bibinfo {year}
  {2006})}\BibitemShut {NoStop}%
\bibitem [{\citenamefont {Pellizzari}\ \emph {et~al.}(1995)\citenamefont
  {Pellizzari}, \citenamefont {Gardiner}, \citenamefont {Cirac},\ and\
  \citenamefont {Zoller}}]{PhysRevLett.75.3788}%
  \BibitemOpen
  \bibfield  {author} {\bibinfo {author} {\bibfnamefont {T.}~\bibnamefont
  {Pellizzari}}, \bibinfo {author} {\bibfnamefont {S.~A.}\ \bibnamefont
  {Gardiner}}, \bibinfo {author} {\bibfnamefont {J.~I.}\ \bibnamefont {Cirac}},
  \ and\ \bibinfo {author} {\bibfnamefont {P.}~\bibnamefont {Zoller}},\ }\href
  {\doibase 10.1103/PhysRevLett.75.3788} {\bibfield  {journal} {\bibinfo
  {journal} {Phys. Rev. Lett.}\ }\textbf {\bibinfo {volume} {75}},\ \bibinfo
  {pages} {3788} (\bibinfo {year} {1995})}\BibitemShut {NoStop}%
\bibitem [{\citenamefont {Duan}\ and\ \citenamefont
  {Kimble}(2004)}]{PhysRevLett.92.127902}%
  \BibitemOpen
  \bibfield  {author} {\bibinfo {author} {\bibfnamefont {L.-M.}\ \bibnamefont
  {Duan}}\ and\ \bibinfo {author} {\bibfnamefont {H.~J.}\ \bibnamefont
  {Kimble}},\ }\href {\doibase 10.1103/PhysRevLett.92.127902} {\bibfield
  {journal} {\bibinfo  {journal} {Phys. Rev. Lett.}\ }\textbf {\bibinfo
  {volume} {92}},\ \bibinfo {pages} {127902} (\bibinfo {year}
  {2004})}\BibitemShut {NoStop}%
\bibitem [{\citenamefont {Duan}\ \emph {et~al.}(2005)\citenamefont {Duan},
  \citenamefont {Wang},\ and\ \citenamefont {Kimble}}]{PhysRevA.72.032333}%
  \BibitemOpen
  \bibfield  {author} {\bibinfo {author} {\bibfnamefont {L.-M.}\ \bibnamefont
  {Duan}}, \bibinfo {author} {\bibfnamefont {B.}~\bibnamefont {Wang}}, \ and\
  \bibinfo {author} {\bibfnamefont {H.~J.}\ \bibnamefont {Kimble}},\ }\href
  {\doibase 10.1103/PhysRevA.72.032333} {\bibfield  {journal} {\bibinfo
  {journal} {Phys. Rev. A}\ }\textbf {\bibinfo {volume} {72}},\ \bibinfo
  {pages} {032333} (\bibinfo {year} {2005})}\BibitemShut {NoStop}%
\bibitem [{\citenamefont {S\o{}rensen}\ and\ \citenamefont
  {M\o{}lmer}(2003)}]{PhysRevLett.91.097905}%
  \BibitemOpen
  \bibfield  {author} {\bibinfo {author} {\bibfnamefont {A.~S.}\ \bibnamefont
  {S\o{}rensen}}\ and\ \bibinfo {author} {\bibfnamefont {K.}~\bibnamefont
  {M\o{}lmer}},\ }\href {\doibase 10.1103/PhysRevLett.91.097905} {\bibfield
  {journal} {\bibinfo  {journal} {Phys. Rev. Lett.}\ }\textbf {\bibinfo
  {volume} {91}},\ \bibinfo {pages} {097905} (\bibinfo {year}
  {2003})}\BibitemShut {NoStop}%
\bibitem [{\citenamefont {Brion}\ \emph {et~al.}(2007)\citenamefont {Brion},
  \citenamefont {M\o{}lmer},\ and\ \citenamefont
  {Saffman}}]{PhysRevLett.99.260501}%
  \BibitemOpen
  \bibfield  {author} {\bibinfo {author} {\bibfnamefont {E.}~\bibnamefont
  {Brion}}, \bibinfo {author} {\bibfnamefont {K.}~\bibnamefont {M\o{}lmer}}, \
  and\ \bibinfo {author} {\bibfnamefont {M.}~\bibnamefont {Saffman}},\ }\href
  {\doibase 10.1103/PhysRevLett.99.260501} {\bibfield  {journal} {\bibinfo
  {journal} {Phys. Rev. Lett.}\ }\textbf {\bibinfo {volume} {99}},\ \bibinfo
  {pages} {260501} (\bibinfo {year} {2007})}\BibitemShut {NoStop}%
\bibitem [{\citenamefont {Fleischhauer}\ \emph {et~al.}(2000)\citenamefont
  {Fleischhauer}, \citenamefont {Yelin},\ and\ \citenamefont
  {Lukin}}]{Fleischhauer2000395}%
  \BibitemOpen
  \bibfield  {author} {\bibinfo {author} {\bibfnamefont {M.}~\bibnamefont
  {Fleischhauer}}, \bibinfo {author} {\bibfnamefont {S.}~\bibnamefont {Yelin}},
  \ and\ \bibinfo {author} {\bibfnamefont {M.}~\bibnamefont {Lukin}},\ }\href
  {\doibase http://dx.doi.org/10.1016/S0030-4018(99)00679-3} {\bibfield
  {journal} {\bibinfo  {journal} {Optics Communications}\ }\textbf {\bibinfo
  {volume} {179}},\ \bibinfo {pages} {395 } (\bibinfo {year}
  {2000})}\BibitemShut {NoStop}%
\bibitem [{\citenamefont {Dicke}(1954)}]{PhysRev.93.99}%
  \BibitemOpen
  \bibfield  {author} {\bibinfo {author} {\bibfnamefont {R.~H.}\ \bibnamefont
  {Dicke}},\ }\href {\doibase 10.1103/PhysRev.93.99} {\bibfield  {journal}
  {\bibinfo  {journal} {Phys. Rev.}\ }\textbf {\bibinfo {volume} {93}},\
  \bibinfo {pages} {99} (\bibinfo {year} {1954})}\BibitemShut {NoStop}%
\bibitem [{\citenamefont {Shahriar}\ \emph
  {et~al.}(2007{\natexlab{a}})\citenamefont {Shahriar}, \citenamefont
  {Pradhan}, \citenamefont {Pati}, \citenamefont {Gopal},\ and\ \citenamefont
  {Salit}}]{Shahriar200794}%
  \BibitemOpen
  \bibfield  {author} {\bibinfo {author} {\bibfnamefont {M.}~\bibnamefont
  {Shahriar}}, \bibinfo {author} {\bibfnamefont {P.}~\bibnamefont {Pradhan}},
  \bibinfo {author} {\bibfnamefont {G.}~\bibnamefont {Pati}}, \bibinfo {author}
  {\bibfnamefont {V.}~\bibnamefont {Gopal}}, \ and\ \bibinfo {author}
  {\bibfnamefont {K.}~\bibnamefont {Salit}},\ }\href {\doibase
  http://dx.doi.org/10.1016/j.optcom.2007.05.057} {\bibfield  {journal}
  {\bibinfo  {journal} {Optics Communications}\ }\textbf {\bibinfo {volume}
  {278}},\ \bibinfo {pages} {94 } (\bibinfo {year}
  {2007}{\natexlab{a}})}\BibitemShut {NoStop}%
\bibitem [{\citenamefont {Shahriar}\ \emph
  {et~al.}(2007{\natexlab{b}})\citenamefont {Shahriar}, \citenamefont {Pati},\
  and\ \citenamefont {Salit}}]{PhysRevA.75.022323}%
  \BibitemOpen
  \bibfield  {author} {\bibinfo {author} {\bibfnamefont {M.~S.}\ \bibnamefont
  {Shahriar}}, \bibinfo {author} {\bibfnamefont {G.~S.}\ \bibnamefont {Pati}},
  \ and\ \bibinfo {author} {\bibfnamefont {K.}~\bibnamefont {Salit}},\ }\href
  {\doibase 10.1103/PhysRevA.75.022323} {\bibfield  {journal} {\bibinfo
  {journal} {Phys. Rev. A}\ }\textbf {\bibinfo {volume} {75}},\ \bibinfo
  {pages} {022323} (\bibinfo {year} {2007}{\natexlab{b}})}\BibitemShut
  {NoStop}%
\bibitem [{\citenamefont {Sarkar}\ \emph {et~al.}()\citenamefont {Sarkar},
  \citenamefont {Kim}, \citenamefont {Fang}, \citenamefont {Tu},\ and\
  \citenamefont {Shahriar}}]{resham}%
  \BibitemOpen
  \bibfield  {author} {\bibinfo {author} {\bibfnamefont {R.}~\bibnamefont
  {Sarkar}}, \bibinfo {author} {\bibfnamefont {M.~E.}\ \bibnamefont {Kim}},
  \bibinfo {author} {\bibfnamefont {R.}~\bibnamefont {Fang}}, \bibinfo {author}
  {\bibfnamefont {Y.}~\bibnamefont {Tu}}, \ and\ \bibinfo {author}
  {\bibfnamefont {S.}~\bibnamefont {Shahriar}},\ }\href@noop {} {\bibinfo
  {journal} {to be submitted for publication}\ }\BibitemShut {NoStop}%
\bibitem [{\citenamefont {Arecchi}\ \emph {et~al.}(1972)\citenamefont
  {Arecchi}, \citenamefont {Courtens}, \citenamefont {Gilmore},\ and\
  \citenamefont {Thomas}}]{PhysRevA.6.2211}%
  \BibitemOpen
\bibfield  {journal} {  }\bibfield  {author} {\bibinfo {author} {\bibfnamefont
  {F.~T.}\ \bibnamefont {Arecchi}}, \bibinfo {author} {\bibfnamefont
  {E.}~\bibnamefont {Courtens}}, \bibinfo {author} {\bibfnamefont
  {R.}~\bibnamefont {Gilmore}}, \ and\ \bibinfo {author} {\bibfnamefont
  {H.}~\bibnamefont {Thomas}},\ }\href {\doibase 10.1103/PhysRevA.6.2211}
  {\bibfield  {journal} {\bibinfo  {journal} {Phys. Rev. A}\ }\textbf {\bibinfo
  {volume} {6}},\ \bibinfo {pages} {2211} (\bibinfo {year} {1972})}\BibitemShut
  {NoStop}%
\bibitem [{\citenamefont {Saffman}\ \emph {et~al.}(2010)\citenamefont
  {Saffman}, \citenamefont {Walker},\ and\ \citenamefont
  {M\o{}lmer}}]{RevModPhys.82.2313}%
  \BibitemOpen
  \bibfield  {author} {\bibinfo {author} {\bibfnamefont {M.}~\bibnamefont
  {Saffman}}, \bibinfo {author} {\bibfnamefont {T.~G.}\ \bibnamefont {Walker}},
  \ and\ \bibinfo {author} {\bibfnamefont {K.}~\bibnamefont {M\o{}lmer}},\
  }\href {\doibase 10.1103/RevModPhys.82.2313} {\bibfield  {journal} {\bibinfo
  {journal} {Rev. Mod. Phys.}\ }\textbf {\bibinfo {volume} {82}},\ \bibinfo
  {pages} {2313} (\bibinfo {year} {2010})}\BibitemShut {NoStop}%
\bibitem [{\citenamefont {Steane}\ \emph {et~al.}(1992)\citenamefont {Steane},
  \citenamefont {Chowdhury},\ and\ \citenamefont {Foot}}]{Steane:92}%
  \BibitemOpen
  \bibfield  {author} {\bibinfo {author} {\bibfnamefont {A.~M.}\ \bibnamefont
  {Steane}}, \bibinfo {author} {\bibfnamefont {M.}~\bibnamefont {Chowdhury}}, \
  and\ \bibinfo {author} {\bibfnamefont {C.~J.}\ \bibnamefont {Foot}},\ }\href
  {\doibase 10.1364/JOSAB.9.002142} {\bibfield  {journal} {\bibinfo  {journal}
  {J. Opt. Soc. Am. B}\ }\textbf {\bibinfo {volume} {9}},\ \bibinfo {pages}
  {2142} (\bibinfo {year} {1992})}\BibitemShut {NoStop}%
\bibitem [{\citenamefont {Urban}\ \emph {et~al.}(2009)\citenamefont {Urban},
  \citenamefont {Johnson}, \citenamefont {Henage}, \citenamefont {Isenhower},
  \citenamefont {Yavuz}, \citenamefont {Walker},\ and\ \citenamefont
  {Saffman}}]{SaffmanNatPhys}%
  \BibitemOpen
  \bibfield  {author} {\bibinfo {author} {\bibfnamefont {E.}~\bibnamefont
  {Urban}}, \bibinfo {author} {\bibfnamefont {T.~A.}\ \bibnamefont {Johnson}},
  \bibinfo {author} {\bibfnamefont {T.}~\bibnamefont {Henage}}, \bibinfo
  {author} {\bibfnamefont {L.}~\bibnamefont {Isenhower}}, \bibinfo {author}
  {\bibfnamefont {D.~D.}\ \bibnamefont {Yavuz}}, \bibinfo {author}
  {\bibfnamefont {T.~G.}\ \bibnamefont {Walker}}, \ and\ \bibinfo {author}
  {\bibfnamefont {M.}~\bibnamefont {Saffman}},\ }\href {\doibase
  10.1038/nphys1178} {\bibfield  {journal} {\bibinfo  {journal} {Nat Phys}\
  }\textbf {\bibinfo {volume} {5}},\ \bibinfo {pages} {110} (\bibinfo {year}
  {2009})}\BibitemShut {NoStop}%
\end{thebibliography}%

\end{document}